\documentclass[11pt,a4paper]{article}

\usepackage[a4paper,margin=2.3cm]{geometry}
\usepackage[T1]{fontenc}
\usepackage{lmodern}

\usepackage{amsmath,amssymb,mathtools}
\usepackage{bm}
\usepackage{slashed}
\usepackage{graphicx}
\usepackage{booktabs}
\usepackage{hyperref}
\usepackage{tikz}
\usepackage{pgfplots}
\usetikzlibrary{patterns}
\pgfplotsset{compat=1.16}
\usepgfplotslibrary{fillbetween}
\usepackage[numbers,sort&compress]{natbib}
\usepackage{chngcntr}
\usepackage{float}
\counterwithout{table}{section}
\counterwithout{figure}{section}

\usepackage{authblk} 

\title{An anomaly-free leptophilic $U(1)'_{\ell}$ completion of the Standard Model}

\author[1]{Seyit Okan Kara\thanks{\href{mailto:seyitokankara@gmail.com}{seyitokankara@gmail.com}}}
\affil[1]{Ni\u{g}de \"Omer Halisdemir University, Bor Vocational School, Ni\u{g}de, T\"urkiye}
\date{} 

\begin{document}
\maketitle

\begin{abstract}
We investigate an anomaly-free, family-universal leptophilic
$U(1)'_\ell$ extension of the Standard Model in which quarks
are neutral under the additional gauge symmetry and mixed
gauge anomalies are canceled by an exotic lepton family
that is vector-like under the Standard Model gauge group.
A complex singlet scalar spontaneously breaks $U(1)'_\ell$,
generating the leptophilic gauge-boson mass and contributions
to the exotic fermion masses. In the primitive integer
charge normalization, symmetry breaking preserves a
residual $Z_6$ gauge symmetry whose $Z_2$ subgroup protects
the lightest exotic fermion. The specified field content
permits Dirac neutrino masses while forbidding Majorana
mass operators for ordinary neutrinos. The exotic neutral
sector contains four mixed Majorana mass eigenstates with
diagonal and off-diagonal couplings to the new gauge boson
$Z_\ell$.

We calculate the exotic fermion spectrum and the tree-level
$Z_\ell$ decay widths into kinematically accessible Majorana
pairs. For $M_{Z_\ell}=900\,\mathrm{GeV}$ and
$g_{\mathrm{eff}}=0.10$, the opening of exotic neutral
channels reduces the combined visible charged-lepton
branching fraction by approximately $10.3\%$ in the first
benchmark. A restricted 5041-point scan over two neutral-sector
mixing parameters yields a relative reduction of
approximately $5.6\%$--$10.8\%$.

We further investigate the corresponding resonance signatures
in $\mu^+\mu^-\to\mu^+\mu^-$ and
$\mu^+\mu^-\to e^+e^-$ using leading-order \textsc{CalcHEP} calculations.
At $\sqrt{s}=900\,\mathrm{GeV}$, the exotic Majorana decay
channels suppress the visible resonance cross sections
by approximately $19.6\%$ and $20.9\%$ for the two benchmarks,
respectively, relative to the closed-channel reference.
The absolute differential cross sections exhibit a similar
suppression, whereas the normalized angular distributions
remain nearly unchanged. We also discuss radiatively
generated gauge--kinetic mixing, complementary precision
observables, and renormalization-group consistency.
These results establish a quantitative connection between
the anomaly-canceling exotic fermion spectrum and visible
leptonic resonance signals, without constituting
experimental exclusion limits.
\end{abstract}

\section{Introduction}
\label{sec:intro}

Neutrino oscillations establish that the minimal Standard Model (SM)
description of lepton masses is incomplete~\cite{PDG2010,PDG2022}.
Additional Abelian gauge symmetries provide a well-motivated framework
for investigating new neutral currents in the lepton sector.
Leptophilic $U(1)'_\ell$ interactions may involve family-universal
lepton charges, $q_e=q_\mu=q_\tau$, or flavor-dependent assignments
such as $L_\mu-L_\tau$~\cite{Heeck:2011}.
This distinction is important for gauge consistency: while certain
lepton-flavor differences are anomaly-free within the SM fermion
spectrum, gauging universal lepton number requires additional
anomaly-canceling states.
Leptophilic gauge bosons have been investigated in global
phenomenological analyses~\cite{Buras:2021btx} and earlier
linear-collider studies~\cite{Kara2011JHEP,Kara2024APPB}.
Their phenomenology therefore depends not only on their interactions
with ordinary leptons but also on the additional particle content
associated with a consistent gauge completion.

The theoretical foundations and phenomenological implications of
gauged baryon and lepton numbers have been extensively explored,
including studies by Fileviez P\'erez and
collaborators~\cite{Fileviez2010,Fileviez2011,Fileviez2018,
Fileviez2019,Fileviez2024,Fileviez2025}.
These investigations emphasize the importance of treating anomaly
cancellation, symmetry breaking, and the exotic fermion spectrum
as interconnected elements of the underlying gauge theory.
Schwaller, Tait, and Vega-Morales (2013)
investigated an anomaly-free gauged-lepton-number
framework containing vector-like leptons and
a Dirac dark-matter candidate protected by an
accidental global symmetry~\cite{Schwaller2013}.
Chang and Ng (2019) considered an alternative
anomaly-free charge assignment with nonuniversal
lepton charges and examined its collider
implications~\cite{ChangNg2019}.
The present construction instead employs family-universal lepton
charges and an extended neutral Majorana sector.
Our central question is how the mass spectrum and mixing of the
anomaly-canceling exotic fermions influence the observable
phenomenology of the new gauge boson.
This connection cannot be fully characterized by a four-lepton
effective interaction alone, since such a description does not
resolve the additional decay channels and their kinematic thresholds.
Direct searches for vector-like leptons provide complementary
experimental information~\cite{ATLAS:2023VLL,CMS:2019VLL},
although their interpretation depends on the relevant mass spectra,
production mechanisms, and decay branching fractions.

We consider an anomaly-free $U(1)'_\ell$ extension in which all
SM leptons carry a common charge, while quarks are neutral under
the additional gauge factor.
The mixed gauge anomalies are canceled by an exotic lepton family
that is vector-like under the SM gauge group.
A complex SM-singlet scalar spontaneously breaks $U(1)'_\ell$,
generating the new gauge-boson mass and the symmetry-breaking
contributions to the exotic fermion masses.
We formulate the anomaly-cancellation conditions in a left-handed
Weyl basis and construct the corresponding renormalizable scalar
and fermion interactions.
In the primitive integer charge normalization, the scalar vacuum
preserves a residual $Z_6$ gauge symmetry, whose $Z_2$ subgroup
protects the lightest exotic fermion against decays exclusively
into SM particles.
Within the specified field content, the same charge assignments
permit Dirac Yukawa masses for ordinary neutrinos while forbidding
their Majorana mass operators.
The exotic neutral sector, by contrast, admits Majorana mass terms
and generally contains four mixed Majorana mass eigenstates,
accompanied by two charged Dirac fermions.
A neutral lightest exotic fermion is consequently a possible
stable-state candidate, although establishing its cosmological
viability requires a dedicated analysis~\cite{Arcadi:2018}.

The principal phenomenological consequence of this construction
arises from the additional decay channels of the leptophilic
gauge boson $Z_\ell$.
We calculate the exotic charged and neutral mass spectra,
the gauge couplings in the fermion mass basis, and the partial
widths for diagonal and off-diagonal Majorana-pair final states.
Two benchmark scenarios, BP1 and BP2, demonstrate how different
Majorana mass and mixing patterns redistribute the exotic
partial widths and modify their threshold behavior.
We further perform a restricted scan of 5041 points in two
neutral-sector mixing parameters, keeping the remaining
reference-benchmark inputs fixed.
For vanishing kinetic mixing, the opening of the exotic neutral
channels reduces the combined visible charged-lepton branching
fraction by approximately $5.6\%$--$10.8\%$ across the
investigated parameter region.
These results establish a quantitative relationship between
the exotic fermion spectrum and the visible decay pattern
of the new mediator.

To investigate the collider implications of this relationship,
we study the processes
$\mu^+\mu^-\to\mu^+\mu^-$ and
$\mu^+\mu^-\to e^+e^-$ at a muon collider.
The two reactions provide complementary information because
elastic muon scattering receives both $s$- and $t$-channel
contributions, whereas electron-pair production proceeds
through $s$-channel exchange at tree level.
Using leading-order \textsc{CalcHEP} calculations~\cite{Belyaev2013,Pukhov2004},
we compare a reference scenario with closed exotic decay
channels against BP1 and BP2.
For $M_{Z_\ell}=900\,\mathrm{GeV}$ and
$g_{\mathrm{eff}}=0.10$, the opening of the exotic
Majorana channels suppresses the visible resonance
cross sections by approximately $20\%$ relative to
the closed-channel reference.
The differential cross sections exhibit corresponding
reductions in magnitude, while their normalized angular
profiles remain nearly unchanged.
These results show that the exotic Majorana sector can
produce substantial modifications of visible leptonic
resonance rates without generating comparably pronounced
changes in the angular shapes.
They also illustrate why different underlying neutral
mass and mixing configurations may yield closely
similar visible collider signatures.

Gauge--kinetic mixing between $U(1)_Y$ and $U(1)'_\ell$
provides an additional phenomenological ingredient~\cite{Holdom:1986}.
For the field content considered here, the mixed fermionic
charge trace is $\mathrm{Tr}(YQ')=-2S$, where
$S=q_e+q_\mu+q_\tau$.
Its nonvanishing value in the universal benchmark implies
that anomaly cancellation does not eliminate radiatively
generated kinetic mixing.
We therefore retain the renormalized mixing parameter
$\varepsilon$ as an independent phenomenological input.
Mixing-induced modifications of precision $Z$ observables
are constrained by electroweak measurements~\cite{ALEPH2006},
while off-shell charged-lepton scattering probes effective
interactions whose leading coefficients scale as
$g_{\mathrm{eff}}^2/M_{Z_\ell}^2$ in the heavy-mediator
limit~\cite{Eichten:1983}.
The running of the additional Abelian gauge coupling
supplies a complementary theoretical consistency condition.
Future circular-collider precision programs~\cite{FCCee2022},
higher-energy linear colliders~\cite{ILC_Snowmass2022},
and lepton-flavor measurements~\cite{BelleII2021}
offer further experimental opportunities.
We identify the relevant parameter dependences without
claiming a combined precision fit or projected
experimental exclusion limits.

The distinctive contribution of this work is thus not the
introduction of gauged lepton number or vector-like exotic
fermions themselves, but the explicit connection between
an anomaly-free gauge completion, its mixed Majorana
spectrum, and measurable leptonic resonance observables.
By combining mass-spectrum calculations, exotic decay
widths, a controlled parameter scan, and two complementary
muon-collider processes, we demonstrate how the
anomaly-canceling sector can influence visible signals
even when its fermionic states are not directly reconstructed.

The paper is organized as follows.
Section~\ref{sec:gauge} presents the gauge structure and
anomaly-cancellation conditions.
Section~\ref{sec:symmetry} develops the symmetry-breaking
mechanism, residual discrete symmetry, and exotic
fermion mass matrices.
Section~\ref{sec:kinmix} discusses gauge--kinetic mixing
and its consequences for neutral-current interactions
and precision observables.
Section~\ref{sec:pheno} examines the exotic lepton
spectrum, Majorana decay widths, benchmark scenarios,
neutral-mixing scan, and complementary phenomenological
constraints.
Section~\ref{sec:muoncollider} investigates the resonance
and angular signatures of the exotic lepton sector
in elastic muon scattering and electron-pair production.
Finally, Section~\ref{sec:conclusions} summarizes
our findings.

\section{Gauge structure and anomaly cancellation}
\label{sec:gauge}

We consider an extension of the Standard Model (SM) gauge group,
\begin{equation}
G_{\rm SM}\times U(1)'_\ell
=SU(3)_C\times SU(2)_L\times U(1)_Y\times U(1)'_\ell,
\end{equation}
where the additional Abelian symmetry acts on leptons, while SM quarks
remain neutral under $U(1)'_\ell$.

We adopt the hypercharge convention $Q=T_3+Y$, with $Y_{\rm PDG}=2Y$.
Before diagonalizing the Abelian kinetic terms, the covariant derivative is
\begin{equation}
D_\mu=\partial_\mu-igW_\mu^aT^a-ig_Y YB_\mu-ig_\ell Q'B'_\mu.
\end{equation}
Gauge--kinetic mixing is discussed in Sec.~\ref{sec:kinmix}.

The physical left- and right-handed SM leptons carry equal
$U(1)'_\ell$ charges $q_\alpha$ ($\alpha=e,\mu,\tau$). In the
left-handed Weyl basis used for anomaly calculations,
\begin{equation}
Q'(L_\alpha)=q_\alpha,\qquad
Q'(e_\alpha^c)=Q'(\nu_\alpha^c)=-q_\alpha.
\label{eq:smcharges}
\end{equation}
Here $e_\alpha^c$ and $\nu_\alpha^c$ are the left-handed charge
conjugates of the right-handed charged leptons and neutrinos. The SM
Higgs doublet is neutral under $U(1)'_\ell$. For the universal benchmark,
\begin{equation}
q_e=q_\mu=q_\tau=a_\ell,
\label{eq:universal}
\end{equation}
the new gauge boson couples vectorially and universally to SM charged
leptons.

Although the additional Abelian charges are vector-like within each
SM lepton family, the SM lepton sector alone has nonzero mixed gauge
anomalies. With
\begin{equation}
S\equiv q_e+q_\mu+q_\tau,
\end{equation}
the anomaly coefficients are
\begin{align}
\mathcal A_{22'}&\equiv[SU(2)_L]^2U(1)'_\ell=\frac S2,\\
\mathcal A_{YY'}&\equiv[U(1)_Y]^2U(1)'_\ell=-\frac S2,\\
\mathcal A_{Y'Y'}&\equiv U(1)_Y[U(1)'_\ell]^2=0,\\
\mathcal A_{\rm grav}&\equiv[\mathrm{grav}]^2U(1)'_\ell=0,\\
\mathcal A_{Y'Y'Y'}&\equiv[U(1)'_\ell]^3=0.
\label{eq:sm_anomalies}
\end{align}
The $[SU(3)_C]^2U(1)'_\ell$ anomaly vanishes because quarks have
zero $U(1)'_\ell$ charge. The right-handed neutrinos ensure that the
SM lepton contributions to the gravitational and cubic anomalies
cancel family by family.

For $S\ne0$, additional fermions are needed to cancel the two mixed
anomalies. We introduce one pair of SM-vector-like lepton doublets
$(L',L'^c)$, one charged-singlet pair $(E',E'^c)$, and one neutral-singlet
pair $(N',N'^c)$. All are written as left-handed Weyl fields. We choose
\begin{equation}
\boxed{\begin{aligned}
Q'(L')=Q'(L'^c)&=-\frac S2,\\
Q'(E')=Q'(E'^c)&=+\frac S2,\\
Q'(N')=Q'(N'^c)&=+\frac S2.
\end{aligned}}
\label{eq:exotic_charges}
\end{equation}
This symmetric charge choice cancels the anomalies and permits
Majorana mass terms in the exotic neutral-singlet sector after
$U(1)'_\ell$ breaking. The complete field content is given in
Table~\ref{tab:charges}.

\begin{table}[t]
\centering
\small
\begin{tabular}{ccccc}
\toprule
Field & $SU(3)_C$ & $SU(2)_L$ & $Y$ & $Q'$\\
\midrule
$L_\alpha$     & $\mathbf{1}$ & $\mathbf{2}$ & $-1/2$ & $q_\alpha$\\
$e_\alpha^c$   & $\mathbf{1}$ & $\mathbf{1}$ & $+1$   & $-q_\alpha$\\
$\nu_\alpha^c$ & $\mathbf{1}$ & $\mathbf{1}$ & $0$    & $-q_\alpha$\\
\midrule
$L'$           & $\mathbf{1}$ & $\mathbf{2}$ & $-1/2$ & $-S/2$\\
$L'^c$         & $\mathbf{1}$ & $\mathbf{2}$ & $+1/2$ & $-S/2$\\
$E'$           & $\mathbf{1}$ & $\mathbf{1}$ & $-1$   & $+S/2$\\
$E'^c$         & $\mathbf{1}$ & $\mathbf{1}$ & $+1$   & $+S/2$\\
$N'$           & $\mathbf{1}$ & $\mathbf{1}$ & $0$    & $+S/2$\\
$N'^c$         & $\mathbf{1}$ & $\mathbf{1}$ & $0$    & $+S/2$\\
\midrule
$H$            & $\mathbf{1}$ & $\mathbf{2}$ & $+1/2$ & $0$\\
$\Phi$         & $\mathbf{1}$ & $\mathbf{1}$ & $0$    & $S$\\
\bottomrule
\end{tabular}
\caption{Gauge quantum numbers of the SM leptons, exotic fermions,
and scalars. All fermions are left-handed Weyl fields, and
$S=q_e+q_\mu+q_\tau$.}
\label{tab:charges}
\end{table}

The mixed non-Abelian anomaly cancels as
\begin{equation}
\mathcal A_{22'}^{\rm total}
=\frac S2+\frac12\bigl[Q'(L')+Q'(L'^c)\bigr]=0.
\end{equation}
The mixed hypercharge anomaly likewise satisfies
\begin{align}
\mathcal A_{YY'}^{\rm total}
={}&-\frac S2+\frac12\bigl[Q'(L')+Q'(L'^c)\bigr]
\nonumber\\
&+Q'(E')+Q'(E'^c)=0.
\end{align}
The remaining coefficients obey
\begin{align}
\mathcal A_{Y'Y'}^{\rm total}&=0,\\
\mathcal A_{\rm grav}^{\rm total}&=0,\\
\mathcal A_{Y'Y'Y'}^{\rm total}&=0.
\end{align}
In particular, the exotic gravitational and cubic contributions are
\begin{equation}
2(-S)+S+S=0
\end{equation}
and
\begin{equation}
4\left(-\frac S2\right)^3+4\left(\frac S2\right)^3=0,
\end{equation}
respectively. The $U(1)_Y[U(1)'_\ell]^2$ coefficient vanishes
because the members of each exotic SM-vector-like pair have opposite
hypercharges and equal $U(1)'_\ell$ charges. Thus all local gauge
and mixed gravitational anomalies cancel.

For the universal assignment $S=3a_\ell$, the exotic charges are
\begin{equation}
Q'(L')=Q'(L'^c)=-\frac{3a_\ell}{2},\qquad
Q'(E')=Q'(E'^c)=Q'(N')=Q'(N'^c)=+\frac{3a_\ell}{2}.
\label{eq:universal_exotics}
\end{equation}
The scalar singlet $\Phi$ has charge $Q'(\Phi)=S$. Its vacuum
expectation value generates the exotic masses through gauge-invariant
renormalizable interactions, including neutral-singlet Majorana terms
proportional to $\Phi^\dagger N'N'$ and
$\Phi^\dagger N'^cN'^c$. The Yukawa interactions and mass matrices
are developed in the next section.

When $S=0$, the SM lepton sector with right-handed neutrinos is
already anomaly-free and no exotic fermions are required for anomaly
cancellation. Familiar examples include $L_\mu-L_\tau$ and
$L_e-L_\mu$.

Anomaly cancellation does not imply the absence of radiative
Abelian gauge--kinetic mixing. In particular, the mixed fermionic
charge trace is not an anomaly coefficient. For the spectrum in
Table~\ref{tab:charges},
\begin{equation}
\mathrm{Tr}(YQ')=-2S.
\label{eq:kinetic_trace}
\end{equation}
The anomaly-free charge assignments established above
provide the basis for the subsequent analysis.
In Sec.~\ref{sec:symmetry}, we examine how these
assignments determine the allowed scalar and fermion
interactions, the resulting mass spectrum, and the
residual discrete symmetry after spontaneous symmetry
breaking. We then investigate their implications for
gauge--kinetic mixing and neutral-current interactions
in Sec.~\ref{sec:kinmix}.

\section{Symmetry breaking and lepton masses}
\label{sec:symmetry}

The spontaneous breaking of $U(1)'_\ell$ generates the mass of the new gauge boson and sets the scale of the symmetry-breaking contributions to the exotic-fermion masses. We consider a minimal scalar sector consisting of the SM Higgs doublet $H$ and a complex SM-singlet scalar $\Phi$ with $U(1)'_\ell$ charge $q_\Phi=S$. For the universal assignment of Sec.~\ref{sec:gauge}, $q_\Phi=3a_\ell$. The most general renormalizable gauge-invariant scalar potential for $S\neq0$ is
\begin{equation}
\begin{aligned}
V(H,\Phi)={}&-\mu_H^2H^\dagger H-\mu_\Phi^2\Phi^\dagger\Phi
+\lambda_H(H^\dagger H)^2\\
&+\lambda_\Phi(\Phi^\dagger\Phi)^2
+\lambda_{H\Phi}(H^\dagger H)(\Phi^\dagger\Phi).
\end{aligned}
\label{eq:scalar_potential}
\end{equation}
We take the vacuum expectation values to be real and write, in unitary gauge for the electroweak symmetry,
\begin{equation}
H=\frac{1}{\sqrt{2}}\begin{pmatrix}0\\v+h_0\end{pmatrix},
\qquad
\Phi=\frac{1}{\sqrt{2}}(v_\Phi+\phi_0+ia),
\end{equation}
where $v\simeq246\,\mathrm{GeV}$ and $v_\Phi>0$. Stationarity at this vacuum gives
\begin{align}
\mu_H^2&=\lambda_Hv^2+\frac{\lambda_{H\Phi}}{2}v_\Phi^2,\\
\mu_\Phi^2&=\lambda_\Phi v_\Phi^2+\frac{\lambda_{H\Phi}}{2}v^2.
\end{align}
Neglecting the small corrections to the neutral gauge-boson spectrum induced by gauge--kinetic mixing, the new gauge boson has mass
\begin{equation}
\boxed{M_{Z_\ell}=g_\ell|q_\Phi|v_\Phi.}
\label{eq:zprime_mass}
\end{equation}
The field $a$ is the Goldstone mode absorbed by the massive $U(1)'_\ell$ gauge boson. The CP-even fluctuations $(h_0,\phi_0)$ have the squared-mass matrix
\begin{equation}
\mathcal M_S^2=\begin{pmatrix}
2\lambda_Hv^2&\lambda_{H\Phi}vv_\Phi\\
\lambda_{H\Phi}vv_\Phi&2\lambda_\Phi v_\Phi^2
\end{pmatrix}.
\label{eq:scalar_mass_matrix}
\end{equation}
With a rotation convention chosen so that the observed Higgs boson is predominantly $h_0$, the mixing angle obeys
\begin{equation}
\tan 2\theta=\frac{2\lambda_{H\Phi}vv_\Phi}
{2\lambda_\Phi v_\Phi^2-2\lambda_Hv^2}.
\label{eq:scalar_mixing}
\end{equation}
For universal rescaling of the SM-like Higgs couplings and in the absence of additional nonstandard Higgs decays, its inclusive signal strengths satisfy $\mu_h\simeq\cos^2\theta$. This approximation is not used when additional scalar or exotic-fermion decay modes are open. The quartic potential is bounded from below if
\begin{equation}
\lambda_H>0,\qquad\lambda_\Phi>0,\qquad
\lambda_{H\Phi}>-2\sqrt{\lambda_H\lambda_\Phi},
\label{eq:scalar_stability}
\end{equation}
where these strict inequalities provide a convenient sufficient condition. The chosen vacuum is a local minimum in the CP-even directions when both eigenvalues of $\mathcal M_S^2$ are positive, equivalently $4\lambda_H\lambda_\Phi>\lambda_{H\Phi}^2$ for positive $v$ and $v_\Phi$; establishing it as the global minimum additionally requires comparing possible competing extrema.

The exotic fermions introduced in Sec.~\ref{sec:gauge} are vector-like under the SM gauge group and chiral under $U(1)'_\ell$. For the charges in Eq.~\eqref{eq:exotic_charges}, their renormalizable Yukawa interactions can be written as
\begin{equation}
\begin{aligned}
-\mathcal L_Y^{\rm exotic}={}&
y_L\Phi L'L'^c+y_E\Phi^\dagger E'E'^c
+y_N\Phi^\dagger N'N'^c\\
&+\kappa_1L'H^\dagger E'^c+\kappa_2L'^cHE'\\
&+\kappa_3L'HN'+\kappa_4L'^cH^\dagger N'^c\\
&+\frac12y_1\Phi^\dagger N'N'
+\frac12y_2\Phi^\dagger N'^cN'^c
+\mathrm{h.c.},
\end{aligned}
\label{eq:complete_yukawa}
\end{equation}
with Lorentz and $SU(2)_L$ contractions understood. Each term is invariant under the complete gauge group. In particular, the symmetric exotic charge assignment permits both neutral-singlet Majorana operators proportional to $\Phi^\dagger$; these operators distinguish the general neutral spectrum from a purely Dirac limit. Here the displayed Lagrangian specifies the exotic-sector mass interactions relevant to the analysis; it is not intended to replace the ordinary SM Yukawa terms.

After symmetry breaking, we define the singlet-induced masses and electroweak mixing parameters by
\begin{equation}
m_F=\frac{y_Fv_\Phi}{\sqrt2}\quad(F=L,E,N),
\qquad
\mu_i=\frac{y_iv_\Phi}{\sqrt2}\quad(i=1,2),
\end{equation}
\begin{equation}
m_i=\frac{\kappa_iv}{\sqrt2}\qquad(i=1,\ldots,4).
\end{equation}
In general these parameters may be complex. The charged exotic sector is described by
\begin{equation}
\boxed{\mathcal M_C=\begin{pmatrix}m_L&m_1\\m_2&m_E\end{pmatrix}.}
\label{eq:charged_mass_matrix}
\end{equation}
Its two nonnegative physical Dirac masses are the singular values of $\mathcal M_C$,
\begin{equation}
U_L^\dagger\mathcal M_CU_R
=\operatorname{diag}(M_{C_1},M_{C_2}),
\end{equation}
where $U_L$ and $U_R$ are unitary matrices.

The neutral exotic sector contains both Dirac-type and Majorana-type mass terms. In the left-handed Weyl basis
\begin{equation}
\Psi_N^T=(L'^0,L'^{c0},N',N'^c),
\end{equation}
the mass Lagrangian is $-\mathcal L_{\rm mass}^{N}=\tfrac12\Psi_N^T\mathcal M_N\Psi_N+\mathrm{h.c.}$, with the symmetric matrix
\begin{equation}
\boxed{\mathcal M_N=\begin{pmatrix}
0&m_L&m_3&0\\
m_L&0&0&m_4\\
m_3&0&\mu_1&m_N\\
0&m_4&m_N&\mu_2
\end{pmatrix}.}
\label{eq:neutral_mass_matrix}
\end{equation}
The four nonnegative physical masses follow from Takagi factorization,
\begin{equation}
U_N^T\mathcal M_NU_N
=\operatorname{diag}(M_{N_1},M_{N_2},M_{N_3},M_{N_4}),
\end{equation}
where $U_N$ is unitary. For generic nonzero Majorana terms, the four mass eigenstates are Majorana fermions. The neutral-current couplings in the mass basis depend on $U_N$ and need not be diagonal, as discussed in the phenomenological analysis.

For the universal charge assignment, $q_\Phi=3a_\ell$ and $g_{\rm eff}=g_\ell a_\ell$. Equation~\eqref{eq:zprime_mass} then implies $v_\Phi=M_{Z_\ell}/(3|g_{\rm eff}|)$, yielding
\begin{equation}
\boxed{|m_F|=\frac{|y_F|}{3\sqrt2}
\frac{M_{Z_\ell}}{|g_{\rm eff}|}.}
\label{eq:exotic_mass_scale}
\end{equation}
Analogous relations hold for $|\mu_{1,2}|$. These expressions relate the symmetry-breaking contributions to the exotic masses to the gauge-boson mass and coupling. The physical fermion masses also depend on the Yukawa couplings and electroweak mixing parameters, so they are not fixed by $M_{Z_\ell}$ and $g_{\rm eff}$ alone.

The symmetry-breaking pattern also leaves an exact discrete gauge subgroup. In the universal case, choosing the primitive integer normalization gives $Q'(L_\alpha)=2$, $Q'(e_\alpha^c)=Q'(\nu_\alpha^c)=-2$, $Q'(L')=Q'(L'^c)=-3$, $Q'(E')=Q'(E'^c)=Q'(N')=Q'(N'^c)=3$, and $Q'(\Phi)=6$. The vacuum expectation value of $\Phi$ therefore preserves a residual $Z_6$ symmetry. Its $Z_2$ subgroup acts trivially on the SM fields and changes the sign of the exotic fermions, forbidding the decay of the lightest exotic fermion exclusively into SM particles. A neutral stable exotic state is obtained when the lightest exotic fermion is electrically neutral. The same residual symmetry forbids Majorana masses for the ordinary right-handed neutrinos within the specified field content, while allowing their gauge-invariant Dirac Yukawa couplings to the SM lepton doublets.

The minimal exotic spectrum thus consists of two charged Dirac fermions and, generically, four neutral Majorana fermions. When the Higgs-induced mixings and neutral-singlet Majorana Yukawa couplings vanish, the charged and neutral mass matrices separate into Dirac sectors. This limiting case is useful for comparison, while the general Majorana-mixed spectrum provides the basis for the exotic decay channels examined below.

\section{Kinetic mixing and precision constraints}
\label{sec:kinmix}

The presence of two Abelian gauge factors permits the renormalizable gauge--kinetic mixing operator
\begin{equation}
\mathcal L_{\rm kin}=
-\frac14 B_{\mu\nu}B^{\mu\nu}
-\frac14 X_{\mu\nu}X^{\mu\nu}
-\frac{\varepsilon}{2}B_{\mu\nu}X^{\mu\nu},
\label{eq:kinetic_lagrangian}
\end{equation}
where $X_\mu$ is the gauge field of $U(1)'_\ell$ and $\varepsilon$ is a renormalized mixing parameter~\cite{Holdom:1986}. Even when $\varepsilon$ is set to zero at a reference scale, fields carrying both hypercharge and $U(1)'_\ell$ charge generally regenerate it through radiative corrections. Anomaly cancellation places no requirement on the mixed charge trace responsible for this running.

In a mass-independent renormalization scheme, the one-loop mixed coefficient is
\begin{equation}
b_{Y\ell}=
\frac23\sum_{\text{Weyl }f}Y_fQ'_f
+\frac13\sum_{\text{complex }s}Y_sQ'_s,
\label{eq:mixed_beta}
\end{equation}
where the sums include gauge-representation multiplicities. For the fields of Table~\ref{tab:charges}, neither scalar contributes: $Q'(H)=0$ and $Y_\Phi=0$. The contributions from the exotic fermions cancel within the pairs of opposite hypercharge, whereas the SM leptons give
\begin{equation}
\sum_fY_fQ'_f=-2S.
\label{eq:mixed_trace}
\end{equation}
Consequently, when all specified fields are active,
\begin{equation}
\boxed{b_{Y\ell}=-\frac43S.}
\label{eq:mixed_beta_result}
\end{equation}
The coefficient is nonzero for the universal leptophilic assignment $S=3a_\ell$.

To leading order in the mixing parameter, the corresponding renormalization-group equation is
\begin{equation}
\mu\frac{d\varepsilon}{d\mu}
=\frac{g_Yg_\ell}{16\pi^2}b_{Y\ell}
+\mathcal O(\varepsilon).
\label{eq:epsilon_rge}
\end{equation}
Approximating the gauge couplings as constant over an interval with unchanged active particle content yields
\begin{equation}
\varepsilon(\mu)=\varepsilon(\Lambda)
+\frac{g_Yg_\ell}{16\pi^2}b_{Y\ell}
\ln\frac{\mu}{\Lambda}
+\Delta\varepsilon_{\rm th},
\label{eq:epsilon_running}
\end{equation}
where $\Delta\varepsilon_{\rm th}$ denotes threshold and matching effects. Across particle thresholds, the appropriate effective-theory coefficient and matching conditions must be used. Thus, the value of $\varepsilon$ at a specified scale depends on its renormalization boundary condition and the spectrum; it is not fixed uniquely by anomaly cancellation.

\subsection{Neutral gauge-boson mixing}

We determine the physical neutral-current couplings to first order in $\varepsilon$. For $|\varepsilon|\ll1$, the field redefinition
\begin{equation}
B_\mu=\widehat B_\mu-\varepsilon\widehat X_\mu,
\qquad
X_\mu=\widehat X_\mu+\mathcal O(\varepsilon^2)
\label{eq:kinetic_redefinition}
\end{equation}
makes the Abelian kinetic terms canonical at this order. The corresponding interactions become
\begin{equation}
\begin{aligned}
\mathcal L_{\rm int}\supset{}&
g_Y\widehat B_\mu J_Y^\mu\\
&+\widehat X_\mu
\left(g_\ell J_\ell^\mu-\varepsilon g_YJ_Y^\mu\right).
\end{aligned}
\label{eq:canonical_currents}
\end{equation}
Here $J_Y^\mu$ and $J_\ell^\mu$ denote the hypercharge and additional Abelian currents, respectively. The Higgs doublet has vanishing $U(1)'_\ell$ charge, so its vacuum expectation value produces no direct mass mixing in the original gauge basis. The canonical field redefinition nevertheless induces mixing between the massive neutral gauge fields after electroweak symmetry breaking.

We define the electroweak neutral combinations by
\begin{equation}
\begin{aligned}
A_\mu^0&=s_WW_\mu^3+c_W\widehat B_\mu,\\
Z_\mu^0&=c_WW_\mu^3-s_W\widehat B_\mu,
\end{aligned}
\end{equation}
with
\begin{equation}
s_W=\frac{g_Y}{\sqrt{g^2+g_Y^2}},
\qquad
c_W=\frac{g}{\sqrt{g^2+g_Y^2}}.
\end{equation}
The photon remains massless. In the $(Z^0,\widehat X)$ basis, the mass-squared matrix reads
\begin{equation}
\mathcal M_V^2=
\begin{pmatrix}
M_{Z,0}^2&\varepsilon s_WM_{Z,0}^2\\
\varepsilon s_WM_{Z,0}^2&M_{X,0}^2
\end{pmatrix}
+\mathcal O(\varepsilon^2),
\label{eq:neutral_mass_matrix_gauge}
\end{equation}
where
\begin{equation}
M_{Z,0}^2=\frac14(g^2+g_Y^2)v^2,
\qquad
M_{X,0}^2=g_\ell^2q_\Phi^2v_\Phi^2.
\end{equation}
The displayed diagonal masses differ from the physical eigenvalues only at second order in the mixing, away from the degenerate region.

For
\begin{equation}
|\varepsilon s_WM_{Z,0}^2|
\ll |M_{X,0}^2-M_{Z,0}^2|,
\end{equation}
and with the sign convention
\begin{equation}
Z_\mu\simeq Z_\mu^0-\xi\widehat X_\mu,
\qquad
Z_{\ell\mu}\simeq\widehat X_\mu+\xi Z_\mu^0,
\label{eq:physical_rotation}
\end{equation}
the mixing angle is
\begin{equation}
\boxed{
\xi\simeq
\frac{\varepsilon s_WM_{Z,0}^2}
{M_{X,0}^2-M_{Z,0}^2}.
}
\label{eq:z_mixing_angle}
\end{equation}
Writing $g_Z=\sqrt{g^2+g_Y^2}$, the predominantly leptophilic state couples to a chiral SM fermion component $f$ with strength
\begin{equation}
\boxed{
g_{Z_\ell f}
=g_\ell Q'_f-\varepsilon g_YY_f
+\xi g_Z(T_3^f-s_W^2Q_f)
}
\label{eq:physical_zprime_coupling}
\end{equation}
up to terms of second order in the mixing. The shift of the SM-like $Z$ coupling is
\begin{equation}
\delta g_{Zf}
=-\xi g_\ell Q'_f
+\mathcal O(\varepsilon^2,\varepsilon\xi,\xi^2).
\label{eq:z_coupling_shift}
\end{equation}
These expressions use the charges of the physical left- or right-handed fermion components, rather than those of their left-handed charge-conjugate fields in Table~\ref{tab:charges}. For quarks, $Q'_q=0$, and the induced $Z_\ell$ interactions arise from the hypercharge-current term and the admixture of the SM neutral current. Hence the gauge boson can contribute to hadronic observables despite the absence of direct $U(1)'_\ell$ charges for quarks.

In the decoupling regime $M_{X,0}^2\gg M_{Z,0}^2$, the mixing angle simplifies to
\begin{equation}
\xi\simeq
\varepsilon s_W\frac{M_{Z,0}^2}{M_{X,0}^2}.
\label{eq:mixing_decoupling}
\end{equation}
The small-angle expansion is inappropriate near $M_{X,0}\simeq M_{Z,0}$; in that region the full neutral mass matrix must be diagonalized. The expressions above consistently retain only the leading effects of kinetic mixing, and they do not constitute a complete electroweak precision fit.

\subsection{Precision observables}

Kinetic and mass mixing affect the couplings of the physical neutral gauge bosons and thereby enter electroweak precision tests. Relevant measurements include the $Z$-pole observables at LEP, low-energy parity-violating processes, and dilepton production at hadron colliders. Their interpretation depends on $M_{Z_\ell}$, the leptonic gauge coupling, the renormalized mixing parameter $\varepsilon$, and the decay channels available to the new mediator. In particular, decays into exotic fermions modify its total width and leptonic branching fractions, which enter resonance-based searches.

We therefore treat $\varepsilon$ as a separate phenomenological parameter alongside $(M_{Z_\ell},g_{\rm eff})$, rather than identifying anomaly cancellation with negligible kinetic mixing. Any exclusion or sensitivity projection displayed solely in the $(M_{Z_\ell},g_{\rm eff})$ plane requires an explicit choice of $\varepsilon$ at a stated scale, together with the assumed exotic spectrum. Without a dedicated combined fit, the discussion here establishes the structure of the relevant precision effects and does not assign a universal numerical bound on $|\varepsilon|$.

\section{Exotic Lepton Spectrum and Phenomenological Implications}
\label{sec:pheno}

The anomaly-free leptophilic $U(1)'_\ell$ framework links neutral-current observables to the spectrum of the exotic fermions required by anomaly cancellation. In the universal benchmark, the gauge boson $Z_\ell$ couples vectorially to the SM charged leptons, while its couplings to quarks are induced by kinetic and neutral gauge-boson mixing, as described in Sec.~\ref{sec:kinmix}. The relevant parameters are the mediator mass $M_{Z_\ell}$, the effective leptonic coupling $g_{\rm eff}=g_\ell a_\ell$, the renormalized kinetic-mixing parameter $\varepsilon$, and the scalar and exotic-fermion masses and mixing matrices. Previous studies have explored complementary collider and flavor probes of leptophilic interactions~\cite{Buras:2021btx}; we examine the exotic spectrum, its consequences for the mediator decay pattern, and complementary precision and consistency probes.

\subsection{Exotic-fermion spectrum and decay channels}

Section~\ref{sec:symmetry} establishes that the universal charge assignment leaves a residual $Z_6$ gauge symmetry whose $Z_2$ subgroup renders the lightest exotic fermion stable within the stated field content. We focus on spectra in which this state is neutral. The charged and neutral fermion mass matrices are diagonalized by the transformations defined in Sec.~\ref{sec:symmetry}; the resulting electroweak currents determine the transitions among mass eigenstates. The gauge-charge protection of the lightest state does not fix its relic abundance or its experimental viability.

Our reference benchmark (BP1) uses $M_{Z_\ell}=900~\mathrm{GeV}$, $g_{\rm eff}=0.10$, and hence $v_\Phi=3~\mathrm{TeV}$, with real mass matrices
\begin{equation}
\mathcal M_C^{\rm BP1}=
\begin{pmatrix}650&80\\90&850\end{pmatrix}
\mathrm{GeV},\qquad
\mathcal M_N^{\rm BP1}=
\begin{pmatrix}
0&650&75&0\\
650&0&0&65\\
75&0&100&350\\
0&65&350&150
\end{pmatrix}\mathrm{GeV}.
\label{eq:bp1_matrices}
\end{equation}
The corresponding physical masses are listed in Table~\ref{tab:exotic_benchmark}. The lightest neutral state is predominantly singlet-like, with an electroweak-doublet fraction of approximately $2.54\%$.

\begin{table}[H]
\centering
\begin{tabular}{cc}
\toprule
State & Mass (GeV)\\
\midrule
$N_1$ & 214.66\\
$N_2$ & 451.39\\
$C_1$ & 618.77\\
$N_3$ & 661.30\\
$N_4$ & 674.57\\
$C_2$ & 881.26\\
\bottomrule
\end{tabular}
\caption{Physical exotic-fermion masses in BP1. The lightest exotic fermion is electrically neutral.}
\label{tab:exotic_benchmark}
\end{table}

The charged-current couplings support the two-body decays $C_i^\pm\to N_1W^\pm$. After removing the common factor $g/\sqrt2$, the magnitudes of the chiral coefficients are $(|G_L^{11}|,|G_R^{11}|)=(0.1000,0.1111)$ and $(|G_L^{21}|,|G_R^{21}|)=(0.0364,0.0413)$. The mass differences are $M_{C_1}-M_{N_1}=404.11~\mathrm{GeV}$ and $M_{C_2}-M_{N_1}=666.60~\mathrm{GeV}$, so both decays are kinematically open. The heavier neutral states also have nonzero off-diagonal SM-$Z$ overlap factors, $(|\mathcal O_{21}^Z|,|\mathcal O_{31}^Z|,|\mathcal O_{41}^Z|)=(0.0527,0.0083,0.1500)$, and all three decays $N_i\to N_1Z$ for $i=2,3,4$ are open. These observations establish an explicit tree-level decay network leading toward the stable neutral state. The individual branching fractions depend on all competing electroweak, scalar, and new-gauge-boson channels. Existing searches for vector-like leptons supply experimental context, with mass limits to be interpreted using the relevant production rates and decay topologies~\cite{ATLAS:2023VLL,CMS:2019VLL}.

\subsection{Leptophilic resonance and Majorana thresholds}

The neutral exotic fermions modify the total $Z_\ell$ width through both diagonal and off-diagonal Majorana-pair channels. In the left-handed basis $(L^{\prime0},L^{\prime c0},N',N'^c)$, and in the charge normalization $a_\ell=1$, define
\begin{equation}
Q_N=\operatorname{diag}(-3/2,-3/2,3/2,3/2),
\qquad G_N=U_N^\dagger Q_NU_N,
\label{eq:neutral_charge_coupling}
\end{equation}
where $U_N^T\mathcal M_NU_N=\operatorname{diag}(M_{N_i})$ is the positive-mass Takagi decomposition. Unitarity implies $G_N^2=(9/4)\mathbf1$, a useful check on the numerical diagonalization. For BP1, representative entries are $G_{11}=1.42389$, $G_{12}=-0.00944i$, $G_{13}=0.47104$, and $G_{14}=0.02378i$ in a specified Takagi phase convention.

At tree level, with $\varepsilon=0$, the partial width for distinct Majorana eigenstates is
\begin{align}
\Gamma_{ij}={}&\frac{g_{\rm eff}^2M_{Z_\ell}}{12\pi}
\sqrt{\lambda(1,x_i,x_j)}
\nonumber\\
&\times\left[
|G_{ij}|^2\left(1-\frac{x_i+x_j}{2}
-\frac{(x_i-x_j)^2}{2}\right)
-3\operatorname{Re}(G_{ij}^2)\sqrt{x_ix_j}
\right],\qquad i\ne j,
\label{eq:majorana_offdiag_width}
\end{align}
where $x_i=M_{N_i}^2/M_{Z_\ell}^2$ and $\lambda(1,x_i,x_j)=(1-x_i-x_j)^2-4x_ix_j$. For identical Majorana fermions, the diagonal coupling is purely axial and the width becomes
\begin{equation}
\Gamma_{ii}=\frac{g_{\rm eff}^2|G_{ii}|^2M_{Z_\ell}}{24\pi}
(1-4x_i)^{3/2},\qquad 2M_{N_i}<M_{Z_\ell}.
\label{eq:majorana_diag_width}
\end{equation}
All widths are taken to vanish below their respective thresholds. These expressions include the identical-particle factor for the diagonal channels.

In BP1, the open neutral channels are $N_1N_1$, $N_1N_2$, $N_1N_3$, and $N_1N_4$; all other neutral pairs and the charged exotic pairs are kinematically closed. Table~\ref{tab:zellell_widths} reports their contributions. The SM-lepton width, neglecting lepton masses and kinetic mixing, is $\Gamma_{\rm SM\,leptons}=1.432394~\mathrm{GeV}$, yielding $\Gamma_{Z_\ell}=1.597358~\mathrm{GeV}$ and a combined charged-lepton branching fraction $0.716197/1.597358=0.4484$. Closing all exotic channels at fixed gauge couplings would give a charged-lepton branching fraction of $0.5$. The opening of the exotic sector therefore reduces this branching fraction by $10.3\%$ relative to that reference, and the idealized on-pole $e^+e^-\to Z_\ell\to\mu^+\mu^-$ rate by approximately $19.6\%$ when the production and decay couplings are unchanged.

\begin{table}[H]
\centering
\begin{tabular}{lc}
\toprule
Channel & BP1 partial width (GeV)\\
\midrule
$N_1N_1$ & $0.164303$\\
$N_1N_2$ & $0.000016$\\
$N_1N_3$ & $0.000625$\\
$N_1N_4$ & $0.000020$\\
\midrule
Total neutral exotic & $0.164964$\\
\bottomrule
\end{tabular}
\caption{Open $Z_\ell$ decays into neutral exotic fermions in BP1, evaluated at tree level for $\varepsilon=0$.}
\label{tab:zellell_widths}
\end{table}

To examine a different mixing regime within the same gauge-charge assignment, we also consider BP2, retaining $M_{Z_\ell}=900~\mathrm{GeV}$ and $g_{\rm eff}=0.10$ and taking
\begin{equation}
\mathcal M_N^{\rm BP2}=
\begin{pmatrix}
0&464.80&-310.02&0\\
464.80&0&0&287.95\\
-310.02&0&74.69&185.34\\
0&287.95&185.34&-84.81
\end{pmatrix}\mathrm{GeV}.
\label{eq:bp2_neutral_matrix}
\end{equation}
The physical neutral masses are $(300.33,313.23,583.12,585.89)~\mathrm{GeV}$. With the illustrative charged mass matrix
\begin{equation}
\mathcal M_C^{\rm BP2}=
\begin{pmatrix}464.80&80\\90&850\end{pmatrix}\mathrm{GeV},
\label{eq:bp2_charged_matrix}
\end{equation}
the lightest charged state remains heavier than $N_1$ and the decay $C_1^\pm\to N_1W^\pm$ is kinematically accessible. The BP2 neutral partial widths are compared with BP1 in Table~\ref{tab:majorana_benchmark_comparison}. In BP2, the $N_1N_3$ and $N_2N_4$ channels contribute appreciably, with thresholds near $883.45$ and $899.12~\mathrm{GeV}$, respectively. The sum of all open neutral exotic widths is $0.178563~\mathrm{GeV}$, giving $\Gamma_{Z_\ell}=1.610957~\mathrm{GeV}$ under the same massless-SM-lepton approximation.

\begin{table}[H]
\centering
\begin{tabular}{lcc}
\toprule
Channel & BP1 (GeV) & BP2 (GeV)\\
\midrule
$N_1N_1$ & $0.164303$ & $0.052478$\\
$N_1N_2$ & $0.000016$ & $0.001089$\\
$N_1N_3$ & $0.000625$ & $0.058415$\\
$N_1N_4$ & $0.000020$ & $0.000213$\\
$N_2N_2$ & closed & $0.052916$\\
$N_2N_3$ & closed & $0.000011$\\
$N_2N_4$ & closed & $0.013440$\\
\midrule
Total & $0.164964$ & $0.178563$\\
\bottomrule
\end{tabular}
\caption{Tree-level $Z_\ell$ partial widths into neutral exotic pairs at $M_{Z_\ell}=900~\mathrm{GeV}$ and $g_{\rm eff}=0.10$, with $\varepsilon=0$. The two benchmarks have the same field content and gauge charges.}
\label{tab:majorana_benchmark_comparison}
\end{table}

The comparison between BP1 and BP2 illustrates two distinct aspects of the exotic neutral sector. Although the total neutral partial widths are relatively similar, their distributions among the available Majorana-pair channels differ substantially. In BP1, the neutral contribution is dominated by the diagonal \(N_1N_1\) channel, whereas BP2 exhibits appreciable contributions from several diagonal and off-diagonal channels. Thus, the total suppression of visible dilepton branching fractions does not uniquely characterize the underlying neutral spectrum. The individual decay thresholds and channel composition provide complementary information about the Majorana mass and mixing structure, even when the gauge charges and the mediator mass and coupling are unchanged.

We illustrate the associated threshold dependence by defining $\Gamma_{\rm run}(E)$ as the sum of the tree-level partial-width expressions evaluated at $M_{Z_\ell}\to E$, including only kinematically open channels, and taking the massless SM-lepton contribution to scale as $E$. Let $\Gamma_0=\Gamma_{\rm run}(M_{Z_\ell})$. The diagnostic ratio of Breit--Wigner denominators is
\begin{equation}
R(E)=
\frac{(E^2-M_{Z_\ell}^2)^2+E^2\Gamma_0^2}
{(E^2-M_{Z_\ell}^2)^2+E^2\Gamma_{\rm run}^2(E)}.
\label{eq:running_width_ratio}
\end{equation}
For BP1 and BP2, respectively, this gives $R(899~\mathrm{GeV})=1.00096$ and $1.00846$, and $R(901~\mathrm{GeV})=0.99904$ and $0.99513$. The comparison isolates the change associated with the energy dependence of the open-channel widths in this prescribed Breit--Wigner approximation. It does not include the dispersive part of the gauge-boson self-energy, SM interference, beam-energy resolution, or a detector-level event-rate calculation. It demonstrates that the neutral Majorana mixing pattern can redistribute exotic partial widths and enhance the near-threshold variation of this diagnostic without changing the underlying gauge structure.

To test whether the reduction of visible leptonic branching fractions is confined to the two benchmark points, we perform a two-parameter scan around BP1. We vary the Higgs-induced neutral mixing masses $m_3$ and $m_4$ independently over $[0,350]~\mathrm{GeV}$, using 71 equally spaced values for each parameter. All other entries of $\mathcal M_N^{\rm BP1}$, the charged matrix $\mathcal M_C^{\rm BP1}$, $M_{Z_\ell}=900~\mathrm{GeV}$, and $g_{\rm eff}=0.10$ are kept fixed. For each of the 5041 points, we recompute the positive neutral masses by Takagi diagonalization, rotate the $U(1)'_\ell$ charge matrix, and sum all kinematically open tree-level Majorana-pair widths. In this restricted scan the lightest exotic fermion remains neutral, and $C_1^\pm\to N_1W^\pm$ is kinematically open throughout. The scan is a controlled test of the resonance-width dependence on neutral mixing, not a global search over the full Yukawa parameter space.

\begin{figure}[H]
\centering
\includegraphics[width=0.85\columnwidth]{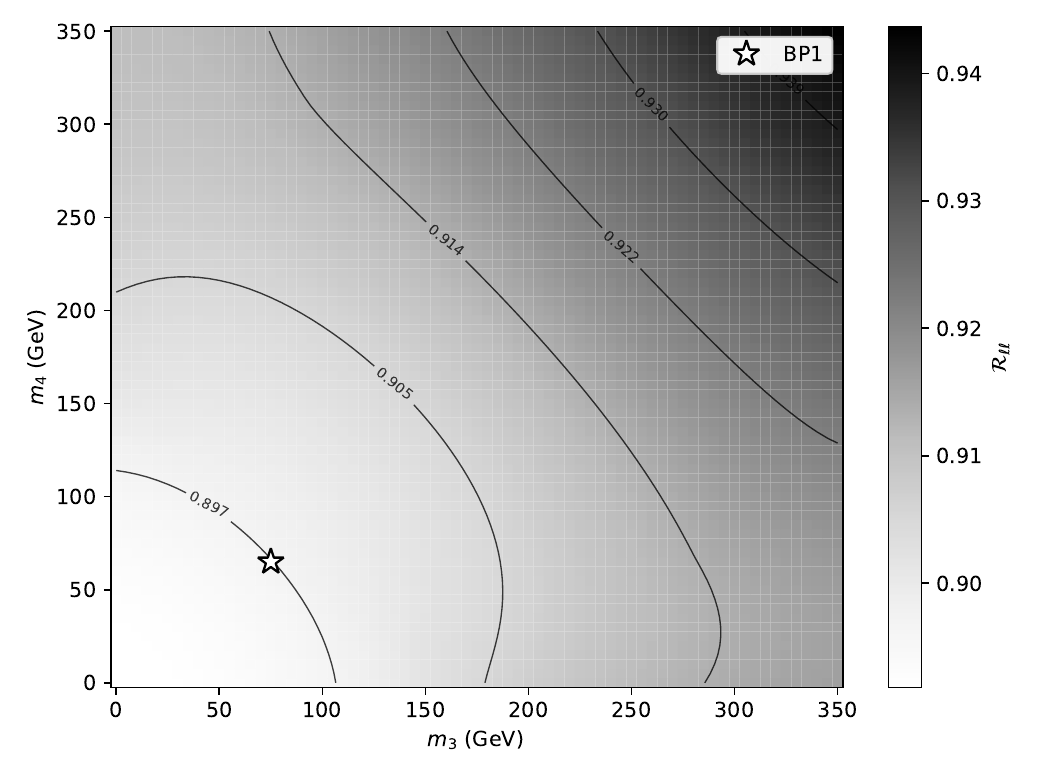}
\caption{Restricted neutral-mixing scan with the BP1 mass parameters fixed except for $m_3,m_4\in[0,350]~\mathrm{GeV}$. The grayscale indicates the ratio $\mathcal R_{\ell\ell}$ of Eq.~\eqref{eq:visible_br_ratio}. The star indicates BP1, $(m_3,m_4)=(75,65)~\mathrm{GeV}$. All 5041 grid points have a neutral lightest exotic fermion and an open $C_1^\pm\to N_1W^\pm$ channel.}
\label{fig:majorana_scan}
\end{figure}

At fixed SM-lepton couplings and for vanishing kinetic mixing, we quantify the change in the combined charged-lepton branching fraction by
\begin{equation}
\mathcal R_{\ell\ell}\equiv
\frac{\mathrm{BR}(Z_\ell\to \ell^+\ell^-)_{\rm exotic\,open}}
{\mathrm{BR}(Z_\ell\to \ell^+\ell^-)_{\rm exotic\,closed}}
=
\frac{\Gamma_{\rm SM\,leptons}}
{\Gamma_{\rm SM\,leptons}+\Gamma_{\rm exotic}},
\label{eq:visible_br_ratio}
\end{equation}
where $\ell^+\ell^-$ denotes the sum over the three charged-lepton flavors in the numerator and denominator. Figure~\ref{fig:majorana_scan} displays the resulting scan. The calculated exotic width ranges from $0.08550$ to $0.17375~\mathrm{GeV}$ and the branching-fraction ratio lies in
\begin{equation}
0.8918\leq\mathcal R_{\ell\ell}\leq0.9437.
\label{eq:visible_br_scan_range}
\end{equation}
Thus, across this specified mixing plane, the opening of the neutral exotic channels reduces the combined charged-lepton branching fraction by approximately $5.6\%$--$10.8\%$ relative to the corresponding closed-channel reference. For an idealized isolated resonance at fixed initial- and final-state couplings, the on-pole dilepton rate scales as $\mathcal R_{\ell\ell}^2$, giving a relative suppression of approximately $10.9\%$--$20.5\%$. These are tree-level width and idealized peak-rate results; they are not collider exclusions. In particular, the scan does not impose precision-electroweak, cosmological, or direct-search constraints on every mass point.
Moreover, scan points with sufficiently light neutral exotic fermions require a separate assessment of constraints from the measured invisible and total \(Z\)-boson widths, accounting for their electroweak-doublet admixtures.

\subsection{Complementary precision probes and phenomenological constraints}
\label{sec:future_circular}

Beyond the on-shell decay patterns, the model is probed by off-shell charged-lepton scattering, precision $Z$ measurements, and mixing-induced interactions with quarks. These observables constrain different combinations of $M_{Z_\ell}$, $g_{\rm eff}$, the renormalized kinetic-mixing parameter $\varepsilon$, and exotic-particle thresholds. The discussion below identifies their parametric dependence without claiming projected experimental exclusions.

\paragraph{Off-shell lepton scattering.}
\label{subsec:contact_norm}
At electron--positron colliders, the additional neutral-current amplitude contributes to $e^+e^-\to\ell^+\ell^-$, with $\ell=e,\mu,\tau$, through interference with photon and SM $Z$ exchange and through its squared contribution. In the heavy-mediator regime $s\ll M_{Z_\ell}^2$, neglecting kinetic mixing and taking $\ell=\mu,\tau$, the leading effective interaction is
\begin{equation}
\mathcal L_{\rm eff}\supset
-\frac{g_{\rm eff}^2}{M_{Z_\ell}^2}
(\bar e\gamma_\mu e)(\bar\ell\gamma^\mu\ell).
\label{eq:lepton_contact}
\end{equation}
We define the corresponding interaction scale by
\begin{equation}
\boxed{\Lambda_\ell\equiv\frac{M_{Z_\ell}}{|g_{\rm eff}|}.}
\label{eq:contact_scale}
\end{equation}
The leading new amplitude relative to the electroweak scattering scale is controlled parametrically by $s/\Lambda_\ell^2$. The attainable sensitivity depends on collision energy, luminosity, beam polarization, angular acceptance, and systematic uncertainties~\cite{Kara2011JHEP,FCCee2022,ILC_Snowmass2022}. The forward--backward asymmetry,
\begin{equation}
A_{\rm FB}=\frac{\sigma_F-\sigma_B}{\sigma_F+\sigma_B},
\label{eq:afb}
\end{equation}
provides complementary information through the angular dependence of the full interference pattern. Its numerical response must be determined from the differential cross section, including the SM vector and axial-vector couplings.

The identical-flavor process $e^+e^-\to e^+e^-$ additionally receives crossed-channel contributions and must be treated using its full scattering amplitude. For $e^+e^-\to\mu^+\mu^-$ or $\tau^+\tau^-$, interference with the SM photon and $Z$ amplitudes can generate leading corrections to suitably chosen observables. Away from points where the interference coefficient vanishes, the relative deviation may be expressed schematically as
\begin{equation}
\frac{\Delta\mathcal O}{\mathcal O_{\rm SM}}
=c_{\mathcal O}(s,\mathrm{cuts})\frac{s}{\Lambda_\ell^2}
+\mathcal O\!\left(\frac{s^2}{\Lambda_\ell^4}\right),
\label{eq:contact_scaling}
\end{equation}
where $c_{\mathcal O}$ depends on the observable, SM electroweak couplings, and analysis acceptance. This expression describes the leading scaling rather than a universal prediction. Total rates and angular observables, including $A_{\rm FB}$ defined in Eq.~\eqref{eq:afb}, probe different combinations of interference and quadratic new-physics contributions. Their relative sensitivities require the differential distribution and experimental uncertainties. The contact approximation itself must be replaced by the full propagator as $\sqrt{s}$ approaches $M_{Z_\ell}$.

\paragraph{$Z$-pole observables and kinetic mixing.}
\label{subsec:Zpole}
At the $Z$ pole, the physical neutral-boson mixing changes the couplings of the observed $Z$. Using the conventions of Sec.~\ref{sec:kinmix}, and away from the nearly degenerate gauge-boson regime, the small mixing angle and leading chiral coupling shift are
\begin{equation}
\xi\simeq
\frac{\varepsilon s_W M_{Z,0}^2}{M_{X,0}^2-M_{Z,0}^2},
\label{eq:xi_small}
\end{equation}
\begin{equation}
\delta g_{Zf}=-\xi g_\ell Q'_f
+\mathcal O(\varepsilon^2,\varepsilon\xi,\xi^2).
\label{eq:gZshift_schematic}
\end{equation}
For a charged-lepton partial width, the leading coupling-induced fractional shift is
\begin{equation}
\frac{\delta\Gamma_\ell}{\Gamma_\ell^{\rm SM}}
\simeq
\frac{2\left(g_{L,\ell}^{\rm SM}\delta g_{L,\ell}
+g_{R,\ell}^{\rm SM}\delta g_{R,\ell}\right)}
{(g_{L,\ell}^{\rm SM})^2+(g_{R,\ell}^{\rm SM})^2},
\label{eq:Zpole_shifts}
\end{equation}
where the expression isolates the vertex-coupling effect and does not include shifts induced through the electroweak input-parameter scheme or additional radiative corrections. Asymmetry observables separately probe the left- and right-handed couplings and therefore provide complementary information.

The nonvanishing mixed trace $\mathrm{Tr}(YQ')=-2S$ obtained in Sec.~\ref{sec:kinmix} implies that kinetic mixing is generally radiatively generated. Its renormalized value depends on the boundary condition and the particle thresholds; anomaly cancellation alone does not determine it. In the decoupling limit $M_{Z_\ell}^2\gg M_Z^2$, one has $\xi\sim\varepsilon s_WM_Z^2/M_{Z_\ell}^2$, while the leading direct shift of the leptonic $Z$ vertex is proportional to $\xi g_{\rm eff}$. A $Z$-pole sensitivity therefore cannot be translated into a unique curve in the $(M_{Z_\ell},|g_{\rm eff}|)$ plane without specifying $\varepsilon$ and the relevant spectrum.

\paragraph{Additional experimental and cosmological constraints.}
Because the SM quarks are neutral under $U(1)'_\ell$, the leading quark couplings of $Z_\ell$ arise through kinetic and neutral gauge-boson mixing. The induced couplings in Eq.~\eqref{eq:physical_zprime_coupling} allow Drell--Yan production, $pp\to Z_\ell\to\ell^+\ell^-$. Its rate depends on the renormalized value of $\varepsilon$, the mediator mass, and the total width, including any open exotic channels. A quantitative hadron-collider exclusion in the $(M_{Z_\ell},g_{\rm eff})$ plane therefore requires a specified mixing prescription and decay spectrum. Independently, the exotic leptons can be pair-produced through their SM electroweak interactions; the resulting charged leptons, weak bosons, and missing-momentum signatures depend on the mass ordering and the mixing matrices. Published vector-like-lepton searches provide relevant comparisons~\cite{ATLAS:2023VLL,CMS:2019VLL}, rather than directly transferable bounds on the benchmarks defined here.

Neutrino--electron scattering and parity-violating measurements provide further probes of mixing-induced effects~\cite{ALEPH2006,PDG2022,Kara2024APPB}. Future measurements at Belle~II, MOLLER, and FCC-ee
may provide complementary sensitivity to leptonic and mixing-induced interactions~\cite{BelleIIPhysicsBook,MOLLER2023,FCCee2022}. A consistent interpretation of these observables must specify $\varepsilon$ and the exotic spectrum. The benchmarks above are illustrative tree-level spectra and decay calculations, not global electroweak fits or experimental sensitivity forecasts.

The residual discrete symmetry identified in Sec.~\ref{sec:symmetry} protects the lightest exotic fermion. When this state is neutral, its cosmological abundance and interactions are determined by the full set of annihilation, coannihilation, and scattering processes. Cosmological and indirect-detection measurements can constrain such scenarios~\cite{Planck2020,FermiLAT2020,IceCube2021,AMS2022}; no relic-density or direct-detection exclusion is imposed here. The scalar sector may also admit a thermal phase transition relevant to gravitational-wave searches, whose prediction requires a finite-temperature analysis~\cite{Caprini2020,LISA2023}. 

The neutrino-mass selection rule is already established in Sec.~\ref{sec:symmetry}: the residual $Z_6$ symmetry permits SM Dirac Yukawa masses while forbidding Majorana mass operators within the specified field content. We do not introduce a seesaw or leptogenesis sector in this minimal construction. Extensions involving additional heavy neutral leptons can produce further collider signatures~\cite{ATLAS2023HNLprompt,CMS:2024HNL_3l,NA622021} and may address leptogenesis~\cite{Davidson2020,Buchmuller2021}; these lie outside the benchmark model analyzed here.

\paragraph{Future circular-collider interpretation.}
\label{subsec:combined_reach}
Future circular $e^+e^-$ facilities, including FCC-ee and CEPC, can combine high-statistics $Z$-pole measurements with charged-lepton scattering at higher energies. Their sensitivity to this model is governed by the distinct parameter dependencies summarized in Table~\ref{tab:precision_summary}. FCC-ee physics studies provide an example of the precision program and its potential observables~\cite{FCCee2022}; a CEPC-specific numerical projection would require its own machine and detector assumptions. In either case, energy stages, luminosities, polarization where available, acceptance, correlated systematics, and theoretical uncertainties must enter a quantitative reach estimate.

\begin{table}[H]
\centering
\small
\renewcommand{\arraystretch}{1.15}
\begin{tabular}{p{0.19\linewidth}p{0.29\linewidth}p{0.36\linewidth}}
\hline
Probe & Leading parameter dependence & Interpretation \\
\hline
Off-shell dileptons &
$g_{\rm eff}^2/M_{Z_\ell}^2$ &
Contact regime $M_{Z_\ell}^2\gg s$; observable-dependent interference.\\
$Z$-pole observables &
$\xi\sim\varepsilon s_W M_Z^2/M_{Z_\ell}^2$;
$\delta g_{Z\ell}\sim-\xi g_{\rm eff}$ &
Requires a renormalized $\varepsilon$ prescription and relevant thresholds.\\
Gauge-coupling running &
$b_{\rm eff}=23$ above all model thresholds &
Theoretical consistency, not a collider exclusion.\\
\hline
\end{tabular}
\caption{Complementary probes of the universal leptophilic $U(1)'_\ell$ model. The entries give parametric dependencies, not numerical projected sensitivities.}
\label{tab:precision_summary}
\end{table}

The $(M_{Z_\ell},|g_{\rm eff}|)$ plane is thus a useful representation of the off-shell interaction scale, while the full precision interpretation also depends on $\varepsilon$ and the exotic-particle thresholds. A combined projected reach requires an observable-level statistical analysis, which is not attempted here. The separation of contact effects, $Z$-pole shifts, and running constraints makes the physical origin of each prospective test explicit.

\paragraph{Right-handed-neutrino cosmology.}
The presence of three light right-handed neutrinos introduces an additional cosmological constraint on the leptophilic gauge sector. If these states were thermalized in the early Universe, their contribution to the effective number of relativistic species depends on their decoupling temperature. In the heavy-mediator regime, the interaction rate scales as \(\Gamma_{\nu_R}\propto (g_\ell/M_{Z_\ell})^4T^5\). For the illustrative benchmark \(M_{Z_\ell}=900\,\mathrm{GeV}\) and \(g_\ell=0.10\), an approximate rate estimate places decoupling near the QCD transition, at a temperature of order \(0.25\,\mathrm{GeV}\). For the family-universal charge normalization
$a_\ell=1$, the coupling used in this estimate,
$g_\ell=0.10$, coincides with the effective
leptonic coupling $g_{\mathrm{eff}}=0.10$
adopted in the collider benchmarks. Assuming three thermalized right-handed neutrinos and \(g_{*s}(T_{\mathrm{dec}})\simeq60\), the entropy-dilution approximation gives \(\Delta N_{\mathrm{eff}}\simeq0.30\), close to the representative cosmological upper bound. This estimate indicates that the benchmark is cosmologically sensitive rather than establishing its exclusion or viability. A precise assessment would require the temperature-dependent interaction rates and entropy degrees of freedom across the QCD transition. The collider results presented here should therefore be interpreted as phenomenological benchmarks, without assuming that every benchmark is cosmologically allowed.

\subsection{Renormalization-group consistency}
\label{subsec:RGE}

The Abelian gauge coupling runs according to
\begin{equation}
\frac{dg_\ell}{d\ln\mu}=\frac{b_\ell}{16\pi^2}g_\ell^3,
\qquad
b_\ell=\frac23\sum_{\rm Weyl\ f}(Q'_f)^2
+\frac13\sum_{\rm complex\ s}(Q'_s)^2.
\label{eq:RGE_gell}
\end{equation}
For the full field content of Table~\ref{tab:charges}, with $a_\ell=1$, the charge-squared sums are $\sum_f(Q'_f)^2=30$ and $\sum_s(Q'_s)^2=9$. They give
\begin{equation}
b_\ell=23.
\label{eq:bell_value}
\end{equation}
Under a common rescaling of all $U(1)'_\ell$ charges, the coefficient for the physical coupling $g_{\rm eff}=a_\ell g_\ell$ remains $b_{\rm eff}=23$. These coefficients describe the regime in which the complete specified particle content contributes to the running; effective-theory coefficients and matching conditions change across mass thresholds.

For a fixed active spectrum between $\mu_0$ and $\mu$, the one-loop solution is
\begin{equation}
\frac{1}{g_{\rm eff}^2(\mu)}
=\frac{1}{g_{\rm eff}^2(\mu_0)}
-\frac{23}{8\pi^2}\ln\frac{\mu}{\mu_0}.
\label{eq:landau_pole}
\end{equation}
Its formal Landau-pole scale is
\begin{equation}
\Lambda_{\rm LP}=\mu_0
\exp\!\left[\frac{8\pi^2}{23g_{\rm eff}^2(\mu_0)}\right].
\label{eq:pert_bound}
\end{equation}
The formal pole indicates the failure of the one-loop perturbative extrapolation. A more restrictive perturbativity requirement may be imposed at a specified ultraviolet scale, once the reference scale and threshold prescription have been fixed. Such a requirement is a theoretical consistency condition and is not an experimental exclusion.

The exotic-lepton spectra and Majorana-pair widths established above thus determine how much of the $Z_\ell$ decay probability remains in visible leptonic final states, while the complementary precision and consistency probes delimit the wider phenomenological setting. A direct consequence of the additional neutral decay channels is a modification of the visible $Z_\ell$ resonance even when the exotic fermions are not reconstructed. In the following section, we quantify this effect in elastic muon scattering and electron-pair production at a muon collider, comparing the closed-channel reference with BP1 and BP2.

\section{Muon-Collider Signatures of the Exotic Lepton Sector}
\label{sec:muoncollider}

The exotic Majorana neutral leptons introduced in the
preceding sections provide additional decay modes for
the leptophilic gauge boson $Z_\ell$, establishing a
direct connection between the structure of the exotic
fermion sector and observable lepton-collider signatures.
When these decay channels become kinematically accessible,
their contributions to the total $Z_\ell$ width modify
the resonance behavior in visible leptonic final states,
even without the direct production or reconstruction
of the exotic particles.

To investigate this connection, we consider two
complementary processes at a muon collider:
elastic muon scattering,
$\mu^+\mu^-\to\mu^+\mu^-$, and electron-pair production,
$\mu^+\mu^-\to e^+e^-$.
Their different exchange topologies provide a useful
framework for examining how the exotic Majorana sector
affects resonance line shapes and angular distributions.
By comparing a reference scenario with two representative
exotic-lepton benchmarks, we identify the observable
consequences of additional decay channels and assess
the extent to which different Majorana mass and mixing
configurations can be distinguished through visible
lepton-pair production.

\subsection{Collider setup and benchmark scenarios}
\label{subsec:muon_setup}

We investigate the resonance signatures of the leptophilic
gauge boson $Z_\ell$ at a muon collider through the processes
\begin{equation}
\mu^+\mu^-\to\mu^+\mu^-,
\qquad
\mu^+\mu^-\to e^+e^-.
\label{eq:muon_processes}
\end{equation}
These two channels provide complementary probes of the
resonance structure. While elastic muon scattering receives
both $s$- and $t$-channel contributions, electron-pair
production proceeds through $s$-channel exchange of the
photon, the SM $Z$ boson, and $Z_\ell$.

The numerical calculations are performed at leading order
using \textsc{CalcHEP}. We adopt the reference parameters
\begin{equation}
M_{Z_\ell}=900\,\mathrm{GeV},
\qquad
g_{\mathrm{eff}}=0.10,
\label{eq:muon_reference}
\end{equation}
corresponding to
\begin{equation}
\alpha_\ell
=\frac{g_{\mathrm{eff}}^2}{4\pi}
\simeq 7.9577\times10^{-4}.
\label{eq:muon_alpha}
\end{equation}
For the numerical analysis, we set the renormalized
kinetic-mixing parameter to $\varepsilon=0$ at the
relevant collider scale, consistently with the
benchmark decay calculations of Sec.~\ref{sec:pheno}.
This choice isolates the effects of the exotic
Majorana decay channels on the visible
$Z_\ell$ resonance.

The center-of-mass energy is scanned over
\begin{equation}
890\,\mathrm{GeV}
\leq \sqrt{s}
\leq 910\,\mathrm{GeV},
\label{eq:muon_energy_scan}
\end{equation}
using 101 equally spaced points. In the \textsc{CalcHEP}
implementation, this corresponds to scanning the
incoming beam momentum parameter $P_{\mathrm{cm}}$
from $445$ to $455\,\mathrm{GeV}$, with
$\sqrt{s}\simeq 2P_{\mathrm{cm}}$ for ultrarelativistic
muon beams.

Both incoming beams are taken to be unpolarized.
A common angular acceptance requirement,
\begin{equation}
|\cos\theta|<0.9,
\label{eq:muon_angular_cut}
\end{equation}
is imposed on the outgoing negatively charged lepton.
The calculations include the tree-level SM and
$Z_\ell$ exchange amplitudes and their interference.
Initial-state radiation and beam-energy smearing are
not included in the present resonance scans.

To isolate the impact of exotic neutral-fermion decays, we consider the two benchmark scenarios BP1 and BP2 introduced in Sec.~5, together with a reference scenario denoted by S0, in which the exotic decay channels are closed while the mediator mass and visible-lepton couplings remain unchanged. The total $Z_\ell$ widths are taken from the decay calculations of Sec.~5, with the S0 width determined solely by the SM leptonic channels. This setup allows us to examine how the different Majorana mass and mixing patterns affect the visible resonance signals through their contributions to the total mediator width.

In the \textsc{CalcHEP} implementation, the exotic Majorana fermions are not included as explicit propagating or external states. Instead, their contributions are incorporated through the independently calculated total width supplied to the $Z_\ell$ propagator. All other numerical settings are kept fixed across the three scenarios, so that their comparison isolates the indirect effect of the exotic decay channels on visible lepton-pair production.

\subsection{Elastic muon scattering}
\label{subsec:elastic}

Elastic muon scattering,
$\mu^+\mu^-\to\mu^+\mu^-$,
receives both $s$- and $t$-channel contributions
from the photon, the SM $Z$ boson, and the
leptophilic gauge boson $Z_\ell$.
The $t$-channel exchange provides a
nonresonant contribution, whereas the
$s$-channel $Z_\ell$ propagator generates
a pronounced enhancement near
$\sqrt{s}=M_{Z_\ell}$.

At the resonance energy
$\sqrt{s}=900\,\mathrm{GeV}$,
the calculated cross sections are
$6915.059\,\mathrm{pb}$,
$5561.522\,\mathrm{pb}$, and
$5468.107\,\mathrm{pb}$
for S0, BP1, and BP2, respectively.
Relative to the reference scenario S0,
the resonance peak is suppressed by
approximately $19.57\%$ in BP1 and
$20.92\%$ in BP2.
This behavior follows from the increase
in the total $Z_\ell$ decay width
when the exotic Majorana channels
become kinematically accessible.

Since the mediator mass and visible-lepton
couplings are identical in all three
scenarios, the observed differences
originate from the modified resonance
propagator rather than changes in the
production vertices.
The resulting energy-dependent cross
sections provide a direct comparison
of the resonance line shapes in the
presence and absence of exotic decay
channels.

\begin{figure}[H]
\centering
\includegraphics[width=0.85\columnwidth]
{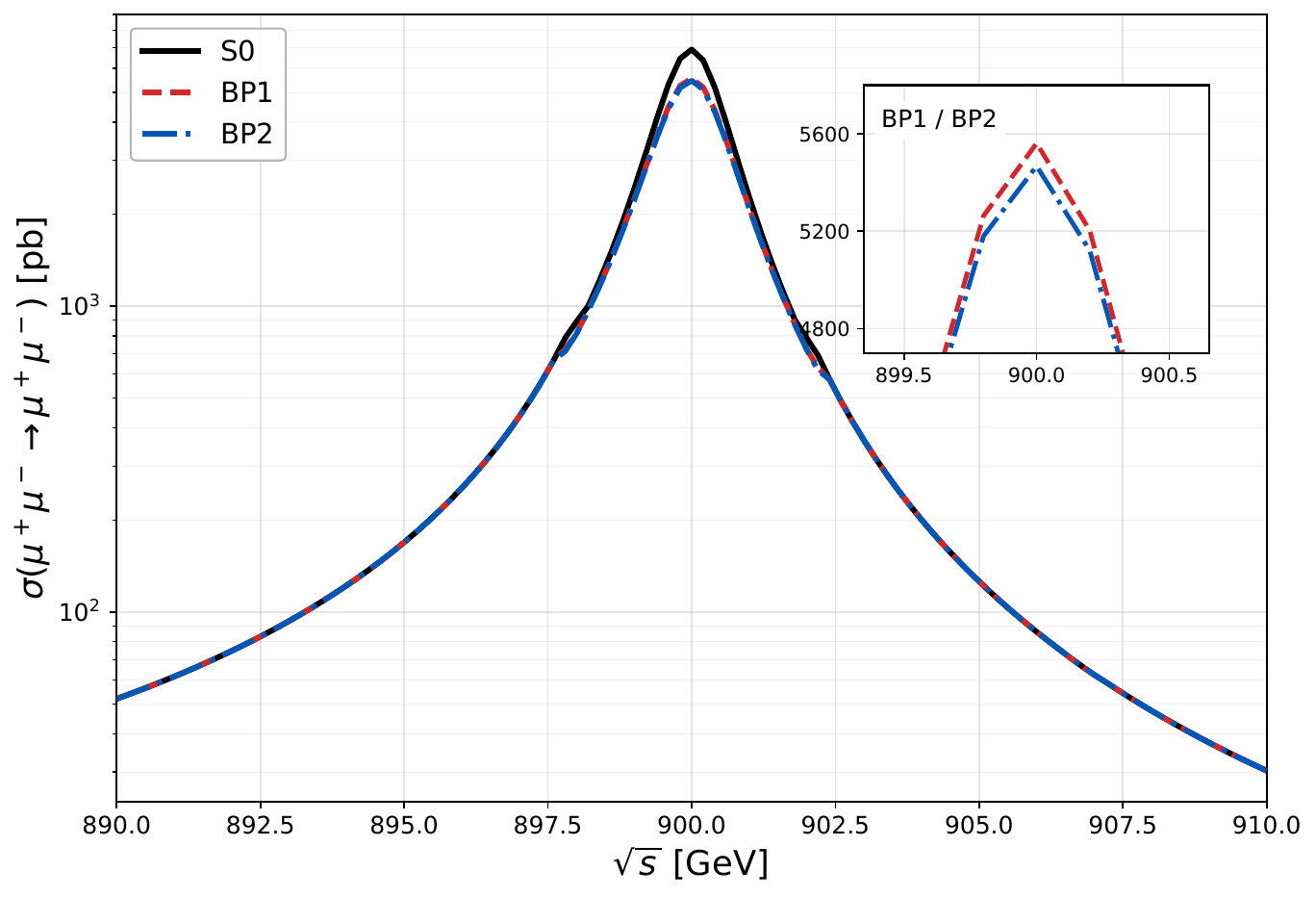}
\caption{
Center-of-mass energy dependence of the
$\mu^+\mu^-\to\mu^+\mu^-$ cross section
for the reference scenario S0 and the
exotic-lepton benchmarks BP1 and BP2.
The calculations are performed with
$M_{Z_\ell}=900\,\mathrm{GeV}$,
$g_{\mathrm{eff}}=0.10$, and
$|\cos\theta|<0.9$.
The opening of exotic Majorana decay channels
suppresses the resonance peak through the
increased total $Z_\ell$ width.
The inset provides an enlarged view of the
BP1 and BP2 resonance profiles, highlighting
their small but distinguishable differences.
}

\label{fig:elastic_resonance}
\end{figure}

The energy dependence of the elastic muon-scattering cross
section is illustrated in Fig.~\ref{fig:elastic_resonance}
for the three benchmark scenarios. The opening of exotic
Majorana decay channels produces a substantial suppression
of the resonance peak relative to S0, whereas the BP1 and
BP2 predictions remain closely spaced. In particular, the
peak cross sections differ by only about $1.68\%$ between
BP1 and BP2, reflecting the small difference in their total
$Z_\ell$ widths. This behavior indicates that the visible
resonance is sensitive to the presence of additional decay
channels but is less effective at distinguishing the detailed
Majorana mass and mixing patterns represented by these
two benchmarks.

\begin{figure}[H]
\centering
\includegraphics[width=0.85\columnwidth]
{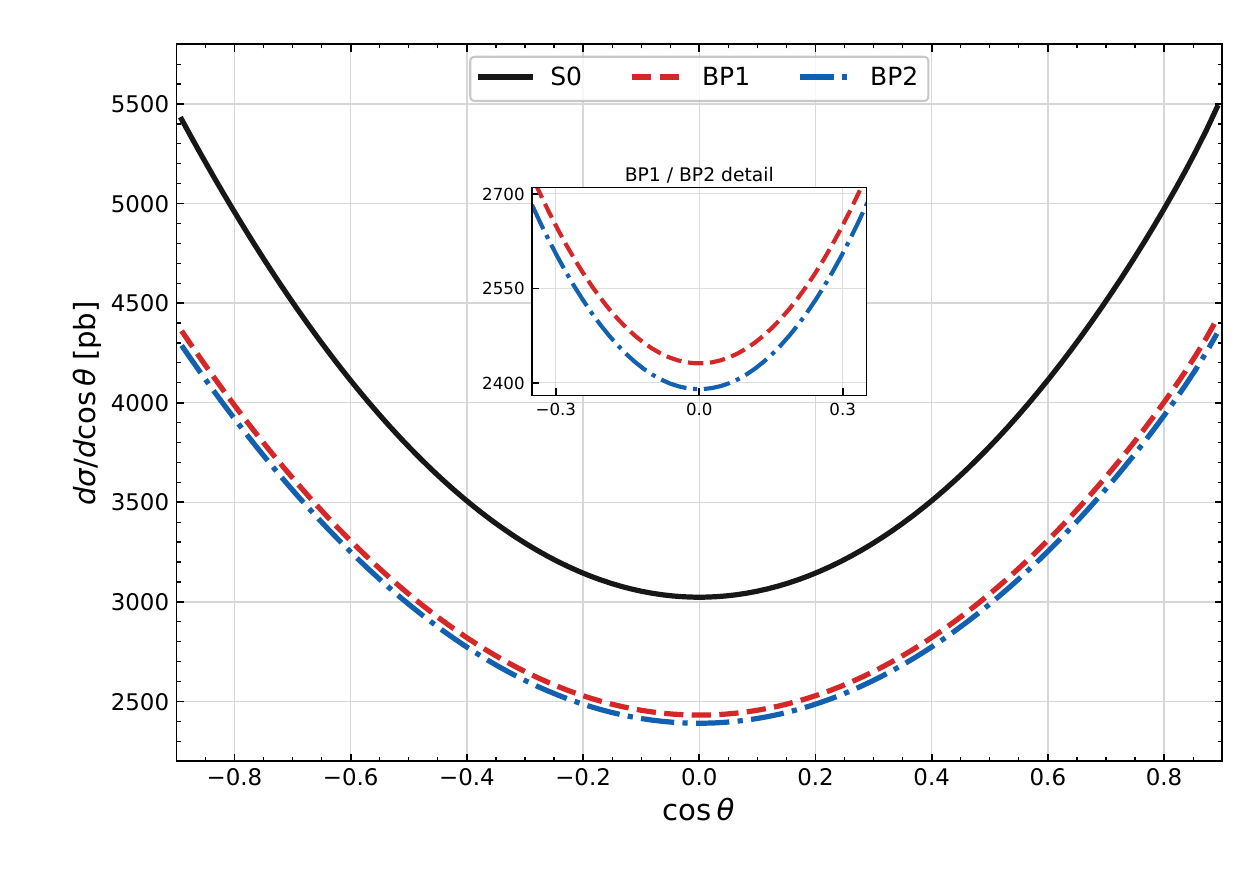}
\caption{
Differential cross section for elastic muon scattering,
$\mu^+\mu^-\to\mu^+\mu^-$, as a function of
$\cos\theta$ for S0, BP1, and BP2 at
$\sqrt{s}=900\,\mathrm{GeV}$.
The calculations are performed with
$M_{Z_\ell}=900\,\mathrm{GeV}$,
$g_{\mathrm{eff}}=0.10$, and
$|\cos\theta|<0.9$.
The inset highlights the small difference between
the BP1 and BP2 predictions.
The opening of exotic Majorana decay channels
reduces the differential cross section while
largely preserving its angular dependence.
}
\label{fig:elastic_angular}
\end{figure}

To further investigate the impact of the exotic
lepton sector, we examine the angular dependence
of elastic muon scattering at the resonance energy,
$\sqrt{s}=900\,\mathrm{GeV}$.
Figure~\ref{fig:elastic_angular} presents the
differential cross sections for the three benchmark
scenarios within the angular acceptance
$|\cos\theta|<0.9$.
All three distributions exhibit a similar angular
profile, with a minimum in the central region
and increasing contributions toward the forward
and backward directions.

The opening of exotic Majorana decay channels
produces a substantial reduction in the magnitude
of the differential cross section, while the
overall angular shape remains nearly unchanged.
Although BP1 and BP2 yield closely spaced
predictions, their small difference is visible
in the enlarged region of the figure.
A comparison of the normalized distributions,
$(1/\sigma)\,d\sigma/d\cos\theta$, shows that
their angular shapes differ by less than
approximately $0.3\%$ over the investigated range.

These results indicate that, for the benchmark
scenarios considered here, the exotic sector
primarily modifies the normalization of the
elastic scattering distribution through the
total $Z_\ell$ width, rather than generating
a significant distortion of its angular shape.
Consequently, the absolute differential cross
section provides a more pronounced signature
of the exotic decay channels than the
normalized angular distribution.

\subsection{Electron-pair production}
\label{subsec:ee}

The electron-pair production process,
$\mu^+\mu^-\to e^+e^-$, proceeds through
$s$-channel exchange of the photon,
the SM $Z$ boson, and the leptophilic
gauge boson $Z_\ell$.
Unlike elastic muon scattering, this process
does not receive a corresponding
$t$-channel contribution at tree level.
Consequently, it provides a particularly
direct probe of the $Z_\ell$ resonance
and its modification by exotic decay channels.

At the resonance energy
$\sqrt{s}=900\,\mathrm{GeV}$,
the calculated cross sections are
$6910.133\,\mathrm{pb}$,
$5556.595\,\mathrm{pb}$, and
$5463.180\,\mathrm{pb}$
for S0, BP1, and BP2, respectively.
Relative to S0, the resonance peak
is suppressed by approximately $19.59\%$
in BP1 and $20.94\%$ in BP2.

The suppression originates from the
increase in the total $Z_\ell$ width
when the exotic Majorana decay channels
become kinematically accessible.
Since the mediator mass and its
couplings to the visible leptons
are held fixed, the differences
among the three scenarios arise
from their distinct total decay widths.

The close agreement between the BP1
and BP2 predictions reflects the
similarity of their total widths,
whereas the substantial separation
from S0 demonstrates the sensitivity
of the visible electron-pair channel
to additional exotic decay modes.

\begin{figure}[H]
\centering
\includegraphics[width=0.85\columnwidth]
{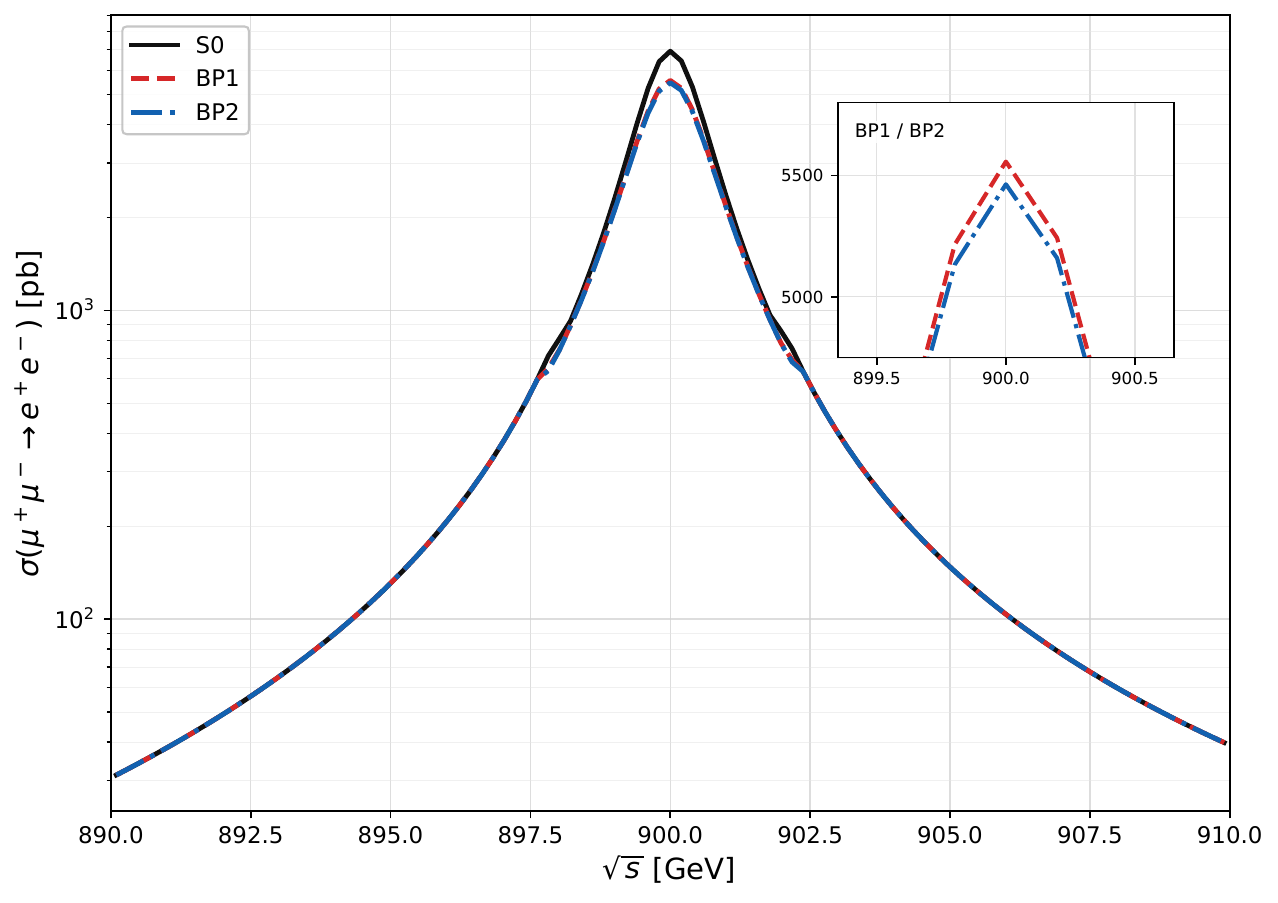}
\caption{
Center-of-mass energy dependence of the
$\mu^+\mu^-\to e^+e^-$ cross section
for the reference scenario S0 and the
exotic-lepton benchmarks BP1 and BP2.
The calculations are performed with
$M_{Z_\ell}=900\,\mathrm{GeV}$,
$g_{\mathrm{eff}}=0.10$, and
$|\cos\theta|<0.9$.
The opening of exotic Majorana decay channels
suppresses the resonance peak through the
increased total $Z_\ell$ width.
The inset provides an enlarged view of the
BP1 and BP2 resonance profiles, highlighting
their small but distinguishable differences.
}
\label{fig:ee_resonance}
\end{figure}

The center-of-mass energy dependence of the
$\mu^+\mu^-\to e^+e^-$ cross section is shown
in Fig.~\ref{fig:ee_resonance} for the three
benchmark scenarios. A pronounced resonance
appears near $\sqrt{s}=M_{Z_\ell}$, with a
substantial reduction in its peak height when
the exotic Majorana decay channels are opened.
The S0 prediction remains clearly separated
from BP1 and BP2 in the resonance region,
whereas the latter two exhibit closely spaced
line shapes. The inset highlights their small
but visible difference near the resonance peak.

Since electron-pair production proceeds entirely
through $s$-channel exchange at tree level,
the observed suppression provides a particularly
transparent manifestation of the increased
total $Z_\ell$ width. The similarity between
BP1 and BP2 further illustrates that distinct
exotic Majorana mass and mixing patterns can
produce nearly indistinguishable visible
resonance signals when their contributions
to the mediator width are comparable.

\begin{figure}[H]
\centering
\includegraphics[width=0.85\columnwidth]
{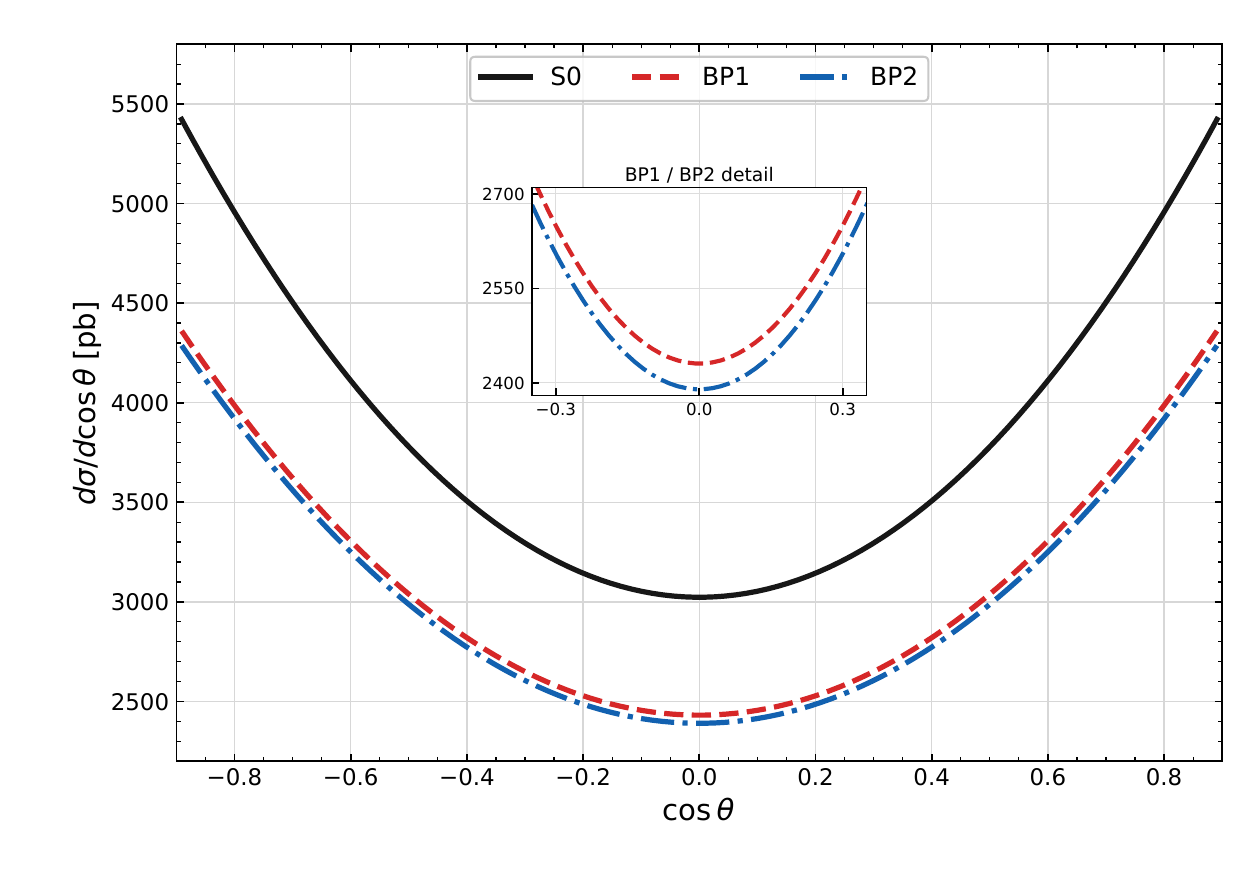}
\caption{
Differential cross section for electron-pair
production, $\mu^+\mu^-\to e^+e^-$, as a
function of $\cos\theta$ for S0, BP1, and BP2
at $\sqrt{s}=900\,\mathrm{GeV}$.
The calculations are performed with
$M_{Z_\ell}=900\,\mathrm{GeV}$,
$g_{\mathrm{eff}}=0.10$, and
$|\cos\theta|<0.9$.
The inset highlights the small difference
between the BP1 and BP2 predictions.
The opening of exotic Majorana decay channels
reduces the differential cross section while
largely preserving its angular dependence.
}
\label{fig:ee_angular}
\end{figure}

To complement the resonance analysis, we examine the
angular dependence of electron-pair production at
$\sqrt{s}=900\,\mathrm{GeV}$.
Figure~\ref{fig:ee_angular} shows the differential
cross sections for S0, BP1, and BP2 within the
acceptance region $|\cos\theta|<0.9$.
All three scenarios exhibit an approximately
symmetric angular distribution, characterized
by a minimum near $\cos\theta=0$ and an
enhancement toward larger absolute values
of $\cos\theta$.

The opening of exotic Majorana decay channels
produces a substantial suppression of the
differential cross section throughout the
investigated angular range, while preserving
the overall angular profile.
The BP1 and BP2 distributions remain closely
spaced, reflecting the similarity of their
total $Z_\ell$ widths.
Their small separation is illustrated in the
inset of Fig.~\ref{fig:ee_angular}.

Unlike elastic muon scattering, electron-pair
production receives no $t$-channel contribution
at tree level.
The observed angular behavior therefore arises
from the $s$-channel exchange amplitudes and
their interference.
Since the visible-lepton couplings are held
fixed across the three scenarios, the
suppression is driven by the modified
$Z_\ell$ propagator through its total width.
This demonstrates that the exotic Majorana
sector can significantly modify the absolute
electron-pair production rate without
introducing a pronounced change in its
angular profile.

\subsection{Comparative resonance analysis}
\label{subsec:comparison}

The resonance analyses of elastic muon scattering
and electron-pair production reveal a common
sensitivity to the exotic Majorana sector,
despite the different diagrammatic structures
of the two processes.
While $\mu^+\mu^-\to\mu^+\mu^-$ receives
both $s$- and $t$-channel contributions,
$\mu^+\mu^-\to e^+e^-$ proceeds exclusively
through $s$-channel exchange at tree level.

Table~\ref{tab:muon_peaks} summarizes the
calculated cross sections at the resonance
energy $\sqrt{s}=900\,\mathrm{GeV}$.
In both channels, the benchmark ordering is
$\sigma_{\mathrm{S0}}>
\sigma_{\mathrm{BP1}}>
\sigma_{\mathrm{BP2}}$.
The opening of exotic Majorana decay channels
reduces the visible resonance rates by
approximately $20\%$ relative to S0.

\begin{table}[H]
\centering
\begin{tabular}{lcc}
\toprule
Scenario
& $\sigma_{\mu\mu}$ (pb)
& $\sigma_{ee}$ (pb) \\
\midrule
S0  & 6915.059 & 6910.133 \\
BP1 & 5561.522 & 5556.595 \\
BP2 & 5468.107 & 5463.180 \\
\bottomrule
\end{tabular}
\caption{
Cross sections at $\sqrt{s}=900\,\mathrm{GeV}$
for the three benchmark scenarios, imposing
$|\cos\theta|<0.9$.
The results are obtained at leading order
without initial-state radiation or
beam-energy smearing.
}
\label{tab:muon_peaks}
\end{table}

Here, $\sigma_{\mu\mu}$ and $\sigma_{ee}$
denote the cross sections for
$\mu^+\mu^-\to\mu^+\mu^-$ and
$\mu^+\mu^-\to e^+e^-$, respectively.
An interesting feature is the close numerical
agreement between the two channels at the
resonance energy.
The difference
$\sigma_{\mu\mu}-\sigma_{ee}$
is approximately $4.93\,\mathrm{pb}$
for all three scenarios, corresponding
to a relative difference below $0.1\%$.
Thus, despite their distinct exchange
topologies, the two processes exhibit
nearly identical integrated resonance
rates for the chosen angular acceptance.

Near an isolated $s$-channel resonance,
the visible cross section approximately
scales as
\begin{equation}
\sigma_{\mathrm{peak}}
\propto
\frac{
\Gamma_{Z_\ell\to\mu^+\mu^-}
\Gamma_{Z_\ell\to f\bar f}
}{
\Gamma_{Z_\ell}^{\,2}
},
\label{eq:resonance_width_scaling}
\end{equation}
when the mediator mass and visible-lepton
couplings are held fixed.
The complete \textsc{CalcHEP} predictions also
include the SM exchange amplitudes
and their interference with $Z_\ell$.
The observed resonance suppression is
therefore consistent with the increased
total mediator width induced by the
additional exotic decay channels.

The angular analysis provides complementary information
on the impact of the exotic Majorana sector. Although
the absolute differential cross sections are substantially
suppressed in BP1 and BP2, their normalized angular
distributions remain nearly unchanged. For elastic muon
scattering, the maximum relative difference between
the normalized distributions is approximately $0.27\%$
over the investigated angular range. The corresponding
differences in electron-pair production are even smaller.
Thus, the exotic decay channels primarily modify the
visible resonance rates rather than the normalized
angular profiles.

Taken together, these results establish a consistent
phenomenological connection between the exotic Majorana
sector and visible lepton-pair production at a muon
collider. Despite the different exchange topologies
of elastic muon scattering and electron-pair production,
both channels exhibit a comparable suppression of their
resonance peaks when additional exotic decay modes
become accessible. At the same time, the angular
distributions retain nearly unchanged profiles,
indicating that the dominant effect of the exotic
sector is transmitted through the total $Z_\ell$
width rather than through modifications of the
visible-lepton interaction structure.

The close agreement between BP1 and BP2 further
illustrates that different Majorana mass and mixing
configurations can lead to similar visible resonance
signatures. Consequently, while the resonance
suppression provides a clear theoretical signature
of additional decay channels within the benchmark
framework considered here, resolving the detailed
structure of the exotic sector would require
complementary information beyond the visible
lepton-pair rates alone.

The benchmark comparison is performed for vanishing
renormalized kinetic mixing at the resonance scale.
For small nonzero values of $\varepsilon$, the
mixing-induced corrections to the visible-lepton
couplings are common to BP1 and BP2 and largely
cancel in their peak-rate ratio in the isolated
$s$-channel approximation. Nevertheless, changes
in the total mediator widths and interference
contributions may affect the quantitative
discrimination between the two benchmarks.

\section{Conclusions}
\label{sec:conclusions}

We have investigated a family-universal leptophilic
$U(1)'_\ell$ extension of the Standard Model in which
quarks carry no direct charge under the additional
gauge symmetry. The mixed gauge anomalies are canceled
by an exotic lepton family that is vector-like under
the Standard Model gauge group but chiral under
$U(1)'_\ell$. Our analysis connects the gauge-consistency
requirements of this construction to the exotic fermion
spectrum, its decay phenomenology, and observable
leptonic resonance signatures.

In the primitive integer charge normalization,
spontaneous symmetry breaking leaves a residual
$Z_6$ discrete gauge symmetry. Its $Z_2$ subgroup
protects the lightest exotic fermion against decays
exclusively into Standard Model particles, allowing
a neutral lightest state to serve as a possible stable
particle candidate. Within the specified field
content, ordinary neutrinos admit Dirac Yukawa masses,
whereas their Majorana mass operators are forbidden.
The exotic neutral sector, in contrast, contains
four mixed Majorana mass eigenstates with diagonal
and off-diagonal couplings to the leptophilic gauge
boson $Z_\ell$, accompanied by two charged Dirac
fermions.

The exotic Majorana spectrum has direct consequences
for the decay structure of $Z_\ell$. We have calculated
the relevant mass eigenstates, rotated gauge couplings,
and partial widths into kinematically accessible
Majorana-pair final states. Two explicit benchmark
scenarios demonstrate how different neutral mass
and mixing configurations redistribute the exotic
partial widths and modify the total mediator width.
For $M_{Z_\ell}=900\,\mathrm{GeV}$ and
$g_{\mathrm{eff}}=0.10$, the combined charged-lepton
branching fraction decreases from $0.5$ in the
closed-channel reference scenario to approximately
$0.4484$ in BP1. A restricted scan of 5041 points
in two neutral-sector mixing parameters yields a
relative suppression of approximately
$5.6\%$--$10.8\%$ in the visible charged-lepton
branching fraction. The corresponding idealized
on-pole dilepton-rate suppression ranges from
approximately $10.9\%$ to $20.5\%$ under the
stated benchmark assumptions.

We have further investigated the collider
manifestations of these exotic decay channels
through elastic muon scattering,
$\mu^+\mu^-\to\mu^+\mu^-$, and electron-pair
production, $\mu^+\mu^-\to e^+e^-$.
Leading-order \textsc{CalcHEP} calculations were performed
for a $900\,\mathrm{GeV}$ leptophilic resonance,
using a common angular acceptance
$|\cos\theta|<0.9$.
At the resonance energy, the opening of exotic
Majorana channels reduces the visible cross
sections by approximately $19.6\%$ in BP1
and $20.9\%$ in BP2 relative to the
closed-channel reference.
Despite their different exchange topologies,
the two processes exhibit closely similar
integrated cross sections at the resonance
energy and comparable suppression patterns.

The angular distributions provide complementary
information about the underlying mechanism.
Although the absolute differential cross sections
are substantially reduced when the exotic channels
are opened, their overall angular profiles remain
nearly unchanged. This behavior reflects the fact
that the visible-lepton couplings are held fixed
across the benchmark scenarios, while the exotic
sector modifies the total $Z_\ell$ width.
Furthermore, BP1 and BP2 exhibit closely spaced
resonance profiles despite their different
Majorana mass and mixing structures.
Thus, visible resonance measurements can be
sensitive to additional exotic decay channels
without necessarily resolving the detailed
composition of the exotic neutral sector.

Complementary phenomenological considerations
include gauge--kinetic mixing, precision
electroweak observables, off-shell leptonic
scattering, and renormalization-group consistency.
For the universal benchmark, the nonvanishing
mixed charge trace,
$\mathrm{Tr}(YQ')=-2S$,
implies that kinetic mixing is generally
radiatively generated and must be specified
through an appropriate renormalization condition.
Precision $Z$-pole observables therefore depend
on additional mixing assumptions, whereas
heavy-mediator charged-lepton scattering probes
the contact-interaction scale
$M_{Z_\ell}/|g_{\mathrm{eff}}|$.
The one-loop running of the new Abelian coupling
provides a separate perturbative-consistency
criterion rather than an experimental exclusion.

Taken together, our results establish a quantitative
connection between the exotic fermions required
for anomaly cancellation and the visible
phenomenology of the leptophilic gauge boson.
The additional Majorana decay channels can
substantially suppress leptonic resonance signals
even when the exotic states are not directly
reconstructed. The combined analysis of decay
widths, resonance line shapes, and angular
distributions therefore provides a coherent
framework for investigating the phenomenological
consequences of this anomaly-free gauge completion.
A complete experimental assessment will require
realistic beam-energy effects, detector-level
analyses, and combined electroweak and collider
constraints. The cosmological viability of a
stable neutral exotic state likewise remains
a separate question.

\bibliographystyle{unsrt} 
\bibliography{refs}

\end{document}